\documentclass[preprint,aps]{revtex4}
\usepackage[dvipdfmx]{graphicx}
\def\vc#1{\mbox{\boldmath $#1$}}
\usepackage[dvipdfmx]{color}

\begin{document}
 
\title{Alpha cluster structures and monopole excitations in $^{13}$C}
 
\author{
	{Taiichi Yamada} and 
	{Yasuro Funaki$^{1}$},
}
 
\affiliation{Laboratory of Physics, Kanto Gakuin University, Yokohama 236-8501, Japan}
  
\affiliation{$^{1}$Nishina Center for Accelerator-based Science, The Institute of Physical and Chemical Research (RIKEN), Wako 351-0098, Japan}

\date{\today}
 
\begin{abstract}
The structure of $1/2^{\pm}$ states in $^{13}$C up to around the $3\alpha+n$ threshold ($E_x = 12.3$~MeV) is investigated with a full four-body $3\alpha+n$ orthogonality condition model (OCM) calculation, where the $3\alpha$ OCM, the model space of which is the subspace of the $3\alpha+n$ model, describes well the structure of the low-lying states of $^{12}$C including the $2^{+}_2$, $0^{+}_3$, and $0^{+}_4$ states, which have been recently observed above the Hoyle state ($0^+_2$).
A full spectrum up to the $1/2^{-}_5$ ($1/2^{+}_3$) state is reproduced consistently with the lowest five $1/2^{-}$  (three $1/2^{+}$) states of experimental spectrum.
It is shown that the $1/2^{-}_2$ and $1/2^{-}_3$ states are characterized by the dominant cluster configurations of $^{9}$Be($3/2^-$,$1/2^-$)+$\alpha$,
while the ground state $1/2^-_1$ has a shell-model-like structure.
The \textcolor{black}{observed} monopole transition strengths to the $1/2^{-}_{2,3}$ states are \textcolor{black}{consistently reproduced for the first time}.
\textcolor{black}{
These results indicate that the excited $1/2^-$ states have cluster structures.
They are compared to those by the previous work with the shell model.
}
On the other hand, the $1/2^{+}_{1}$ state is found to have a loosely bound neutron structure in which the extra neutron moves around $^{12}$C(g.s) core with $1S$ orbit, reflecting the fact that this state appears by $1.9$~MeV just below the $^{12}$C(g.s)+$n$ threshold, while the $1/2^{+}_{2}$ and $1/2^{+}_{3}$ states are characterized by $^{9}$Be+$\alpha$ structures.
We found that the $1/2^+_5$ state located above the $3\alpha+n$ threshold is the Hoyle analogue state in $^{13}$C,  the wave function of which is described by product states of constituent clusters, $(0S)^3_{\alpha}(S)_n$, with the probability of $52$~\%.
\end{abstract}
 
\maketitle
 
\section{Introduction}\label{sec:introduction}

Cluster as well as mean field pictures are important to understand the structure of light nuclei~\cite{wildermuth77,ikeda80,ptp_supple_68,ptp_supple_68_chp3}. 
A lot of cluster states are known to exist in light nuclei~\cite{ptp_supple_68} as well as neutron rich nuclei~\cite{oertzen06,horiuchi12} and hypernuclei~\cite{hiyama09}.
The typical cluster state is the Hoyle state, the second $0^+$ state ($0^+_2$) at $E_x=7.65$ MeV in $^{12}$C.
This state is located just above the $3\alpha$ disintegrated threshold, and is characterized by the large monopole transition rate sharing about 16~\% of the energy-weighted sum rule~\cite{yamada08_monopole}.
The microscopic and semi-microscopic cluster models in 1970's~\cite{horiuchi74,fukushima78,uegaki77} demonstrated that the Hoyle state has a loosely coupled $3\alpha$ cluster structure. 
In 2000's, however, it was found that the Hoyle state has a remarkable aspect of the $\alpha$-particle condensate structure, described as a dominant product state of $\alpha$ particles, all occupying an identical $0S$ orbit with $70~\%$ probability~\cite{tohsaki01,funaki03,matsumura04,yamada05,chernykh07,chernykh10,funaki10,yamada12_review}.    
This has aroused a great interest in nuclear cluster physics, and brought significant developments in experimental studies, in particular, for the excited states of the Hoyle states.
Recent experimental efforts~\cite{itoh04,freer09,itoh11,fynbo11,zimmerman11,zimmerman13} eventually confirmed the second $2^+$ state ($2^+_2$) of $^{12}$C, which had been predicted at a few MeV above the Hoyle state by the microscopic cluster models in 1970's.    
A new observation of the $4^+$ state at $13.3$ MeV was reported in Ref.~\cite{freer11}.
In addition, two broad $0^+$ states, $0^+_3$ and $0^+_4$, were observed~\cite{itoh11} at $9.04$ MeV and $10.56$ MeV with the widths of 1.45 MeV and 1.42 MeV, respectively, the results of which are consistent with the theoretical predictions by the semi-microscopic cluster model, $3\alpha$ OCM (orthogonality condition model), using the complex scaling method~\cite{kurokawa05,ohtsubo13}.
The underlying structure of these newly observed states has been also discussed by the rigid $3\alpha$ cluster picture~\cite{bijker00,marin14} and the container picture~\cite{funaki14}.
According to the experimental and theoretical analyses~\cite{kurokawa05,itoh11,funaki14}, the $0^+_3$ state has a prominent $^8$Be($0^+$)+$\alpha$ structure with a higher nodal behavior, while the $0^+_4$ state is characterized by a linear-chain-like structure having the dominant configuration of ${^8}{\rm Be}(2^+)$+$\alpha$ with a relative $D$-wave motion. 

The discovery of the $\alpha$-particle condensate aspect in the Hoyle state has triggered searching for Hoyle-analog states in other light nuclei such as $^{16}$O~~\cite{funaki08,yamada12}, $^{11}$B~~\cite{kawabata07,enyo07,yamada10}, $^{13}$C~\cite{kawabata08,yamada08_13c}, and $^{56}$Ni~~\cite{akimune13} etc.
The definition of the Hoyle-analog states is that the constituent clusters make nonlocalized motion occupying mainly the lowest orbit (or some excited orbits) of the cluster mean-field potential in nucleus.
Reflecting their gas-like properties, they seem to appear around the disintegrated threshold of the constituent clusters. 
The structure study of $^{16}$O has recently made a significant progress~\cite{funaki08,yamada12,horiuchi14}, following the study of the $^{12}$C+$\alpha$ cluster model in 1970's~\cite{suzuki76} and 1980's~\cite{descouvemont87}.
The six lowest $0^+$ states of $^{16}$O (including the ground state with a doubly closed shell structure) up to $E_{x} \simeq 16$~MeV around the $4\alpha$ threshold were for the first time reproduced well together with the monopole strengths and decay widths within the framework of the $4\alpha$ OCM~\cite{funaki08}.
The OCM is a semi-microscopic cluster model, which is an approximation of RGM (resonating group method) and is extensively described in Ref.~\cite{saito68}. 
Many successful applications of OCM to ordinary nuclei as well as hypernuclei are reported in Refs.~\cite{ptp_supple_68,ptp_supple_68_chp3,hiyama09}. 
According to the $4\alpha$ OCM calculation, the five lowest excited $0^+$ states were found to have cluster structures with the substantial monopole strengths comparable to the single particle monopole strength~\cite{yamada08_monopole,yamada12}.
The $0^{+}_{6}$ state at $15.1$ MeV just above the $4\alpha$ threshold was assigned as a strong candidate of the $4\alpha$ condensate~\cite{yamada12_review}, $(0S)^{4}_{\alpha}$, with the probability of $61~\%$.
A novel structure of $^{12}$C(Hoyle)+$\alpha$ has been discussed in the highly excited energy region around $E_{x} \simeq 18$ MeV by the potential model~\cite{ohkubo10}, the $4\alpha$ OCM~\cite{funaki12}, and the $^{12}$C($0^+_2$)+$\alpha$ model~\cite{dufour14}. 
On the other hand, the $1/2^+_2$ state of $^{11}$B just above the $2\alpha+t$ threshold was pointed out to have a Hoyle analogue structure, $(0S)^2_{\alpha}(0S)_{t}$, with the probability of $60~\%$~\cite{yamada10}.
In addition, multi-$\alpha$ gaslike states~\cite{tohsaki01,tohsaki04,yamada04} have been experimentally explored in $^{36}$Ar, $^{40}$Ca and $^{56}$Ni~\cite{akimune13}. 
An interesting viewpoint of clustering, a container picture, has quite recently been proposed to understand cluster dynamics not only for the gaslike cluster states but also for the cluster states which had been regarded as having localized cluster structures like the inversion doublet band in $^{20}$Ne~\cite{zhou13,zhou14} and linear-chain states~\cite{suhara14}.

Isoscalar (IS) monopole transition strengths are very useful to search for cluster states in the low-energy region~\cite{kawabata07,yamada08_monopole,yamada12}.
The IS monopole excitations to cluster states in light nuclei are in general strong as to be comparable with the single particle strength, and their experimental strengths share about 20\% of the sum rule value in the case of $^{4}$He, $^{11}$B, $^{12}$C, and $^{16}$O etc.~\cite{yamada08_monopole,yamada12}. 
As a typical case in light nuclei, the IS monopole strength function of $^{16}$O~\cite{lui01} was discussed up to $E_{x} \sim 40$ MeV~\cite{yamada12}.
It was found that 1) two different types of monopole excitations exist in $^{16}$O; one is the monopole excitation to cluster states which is dominant in the lower energy part, and the other is the monopole excitation of the mean-field type such as one-particle one-hole ($1p1h$) which is attributed mainly to the higher energy part, and 2) this character of the monopole excitations originates from the fact that the ground state of $^{16}$O with the dominant doubly closed shell structure has a duality of the mean-field-type as well as alpha-clustering character.
According to the Bayman-Bohr theorem~\cite{bayman58}, the dual nature of the ground state seems to be a common feature in light nuclei.
The $4\alpha$ OCM calculation showed that the fine structures at the lower energy region up to $E_{x} \simeq 16$ MeV observed in the experimental IS monopole strength function~\cite{lui01,yamada12} surely correspond to cluster states and are rather satisfactorily reproduced by its calculation~\cite{yamada12}, although the fine structures are much difficult to be reproduced by the mean-field calculations such as RPA and QRPA~\cite{gambacurta10} including the relativistic approach~\cite{ma97} together with beyond mean-field calculations~\cite{bender03}.
These results were supported by a recent microscopic calculation with an extended $^{12}$C+$\alpha$ cluster model~\cite{kanada-enyo14} based on the antisymmetrized molecular dynamics (AMD).
\textcolor{black}{
On the other hand, the recent experiments on the inelastic form factor to the Hoyle state ($0^+_2$) as well as its monopole matrix element by the $^{12}$C$(e,e')$ reactions~\cite{chernykh07,chernykh10} have confirmed the old experimental data~\cite{ajzenberg93}.
Their data are known to be well reproduced by the $3\alpha$ cluster-model calculations~\cite{fukushima78,funaki03} as well as the fermionic molecular dynamics (FMD) approach~\cite{roth04}.
This fact shows that the $\alpha$ cluster model favorably compares with the FMD calculation in describing the monopole transitions, although the model space of the $3\alpha$ cluster model is limited compared with that of FMD. 
}

The IS monopole transition strengths are also useful to search for cluster states in neutron rich nuclei.
Quite recently the enhanced monopole strengths in $^{12}$Be, predicted by the generalized two-center \textcolor{black}{cluster} model (GTCM)~\cite{ito08,ito12,ito14}, 
have been observed in the breakup-reaction experiment using a $^{12}$Be beam~\cite{yang14}.
According to their results, the enhanced monopole state observed corresponds to the $0^{+}$ state at $E_x=10.3$~MeV with an $\alpha$+$^{8}$He cluster structure.   
Thus the IS monopole transition strengths indicate to be a good physical quantity to explore cluster states in light nuclei.

The purpose of the present paper is to study the structure of $^{13}$C from the viewpoint of $\alpha$ clustering, in particular, paying our attention to the $1/2^{-}$ states together with the $1/2^{+}$ ones up to around the $3\alpha+n$ threshold ($E_{x} \sim 12$ MeV).
There exist five $1/2^-$ states ($E_x=0.0$, 8.86, 11.08, 12.5~\cite{kawabata08}, and 14.39~MeV) and three $1/2^+$ states (3.09, 11.00, and 12.14 MeV) up to $E_{x} \sim 12 $ MeV~\cite{ajzenberg93}, where several cluster disintegration thresholds are open: $^{12}$C(g.s)+n ($E_{x}=4.95$ MeV), $^{12}$C(Hoyle)+$n$ ($12.60$), $^{9}$Be(g.s)+$\alpha$ ($10.68$), $^{8}$Be+$^{5}$He ($13.2$), and $3\alpha+n$ ($13.22$) etc. 
The motivations why we focus on studying the structure of the $1/2^{\pm}$ states are threefold.  
The first motivation is originating from the results by the shell model calculations.
The shell model calculation by Millener et al.~\cite{millener89} and the no-core shell model one by Navr\'atil et al.~\cite{navratil07} reasonably reproduced the energy levels of the lowest three $1/2^-$ states and the lowest two $1/2^-$ states, respectively. 
\textcolor{black}{
The $1/2^{-}_{1,2}$ states are interpreted as $p$-shell levels (the dominant configurations are SU(3)$[f](\lambda,\mu)=[441](0,3)$ and $[432](1,1)$, respectively)~\cite{millener89}, based on the experiments with one-nucleon and two-nucleon pickup reactions~\cite{fleming68,hinterberger68} and the $0\hbar\omega$ shell-model calculation~\cite{cohen65} in 1960's, while the $1/2^-_3$ state is assigned as the $2\hbar\omega$ shell-model state~\cite{millener89}.
} 
However, the {\rm C0} transition matrix elements to the $1/2^{-}_{2,3}$ states measured by the \textcolor{black}{$^{13}$C}$(e,e')$ experiments~\cite{wittwer69}, which are the same order as that of the Hoyle state~\cite{chernykh07,chernykh10,ajzenberg93}, are much difficult to be reproduced within their shell-model framework~\cite{millener89}, where {\rm C0} denotes a longitudinal electric {\it monopole} transition. 
According to their results, the difficulty with the transition to the $1/2^{-}_{2}$ $8.86$~MeV state, in particular, is that the initial and final $p$-shell wave functions have different intrinsic spins, $S=1/2$ and $S=3/2$, and thus the calculated {\rm C0} form factor is much smaller than the observed, due to the characteristic of the {\rm C0} transition, i.e.~conserving the intrinsic spin.
As for the $1/2^-_3$ $11.08$~MeV state, the calculated form factor is also much too small, although \textcolor{black}{its state is described by the $2\hbar\omega$ shell-model wave function with} dominant intrinsic spin $S=1/2$.
\textcolor{black}{
One of the possibilities to solve these difficulties in the shell-model approach may be to take into account the mixture of the $(0+2)\hbar\omega$ configurations in each $1/2^{-}$ levels including the ground state.
However, there are no papers reproducing the experimental {\rm C0} matrix elements with the $(0+2)\hbar\omega$ shell-model calculations as far as we know.
}

In addition to the experimental {\rm C0} matrix elements, the IS monopole transition rates of $^{13}$C for the lowest three excited $1/2^-$ states have been reported with the inelastic $\alpha$ scattering on the target of $^{13}$C by Kawabata et al.~\cite{kawabata08}, and their experimental values are comparable to the single particle one~\cite{yamada08_monopole}. 
In the light of the fact that the monopole strengths are one of the good physical quantities to explore cluster states in light nuclei as mentioned above, those experimental results suggest that the low-lying excited $1/2^-$ states have an aspect of $\alpha$-clustering in $^{13}$C.  
Candidates for the molecular states with $K^{\pi}=3/2^{\pm}$~\cite{milin02,freer11_13c,wheldon12} and linear-chain configurations of three $\alpha$ clusters in $^{13}$C~\cite{milin02,itagaki06,furutachi11,suhara11} have been proposed  in highly excited energy region above the $3\alpha+n$ threshold.
However, the cluster structures of the low-lying states including the $1/2^{\pm}$ ones up to around the $3\alpha+n$ threshold have not been studied well, although some preliminary results have been reported~\cite{milin02,yamada08_13c,yoshida09}.
Thus, it is important to study the \textcolor{black}{$\alpha$-clustering aspects in the $1/2^-$ states of} $^{13}$C, focusing on the {\rm C0} transitions and IS monopole transitions.

The second motivation in the present study is related to the recent progress of the structure study of $^{12}$C as mentioned above, which has disclosed richness in the structure of $^{12}$C.  
Thus it is very intriguing to study what kinds of structure changes happen in $^{13}$C when an extra neutron is added into $^{12}$C, which has the shell-model-like ($0^+_1$), $3\alpha$-gas-like ($0^+_2$), higher-nodal $^8$Be($0^+$)+$\alpha$ cluster ($0^+_3$), and linear-chain-like ($0^+_4$) structures etc.
\textcolor{black}{
As mentioned above, the $3\alpha$ cluster models~\cite{kurokawa05,ohtsubo13,funaki14} are known to describe nicely the structures of the $0^+_1$, $0^+_2$, $0^+_3$, and $0^+_4$ states of $^{12}$C (see also Fig.~\ref{fig:2} in this text).
}
On the other hand, the third motivation is to explore the Hoyle-analogue states in $^{13}$C, the wave function of which is described by product states of constituent clusters such as $(0S)^3_{\alpha}(S)_{n}$ etc.
They might appear around the $3\alpha+n$ threshold.

In the present study, we take a four-body $3\alpha+n$ OCM with the Gaussian expansion method (GEM)~\cite{kamimura88,hiyama03}, the model space of which is large enough to cover the $3\alpha+n$ gas, the $^{12}$C+$n$ cluster, $^{9}$Be+$\alpha$ cluster, and $^{8}$Be+$^{5}$He cluster configurations as well as the shell-model configurations. 
It should be reminded that the $3\alpha$ OCM calculation with the complex scaling method~\cite{kurokawa05,
ohtsubo13} succeeded in reproducing well the structure of the low-lying structure up to $E_x\sim 12$~MeV including the recently observed $0^{+}_{3}$, $0^{+}_{4}$, and $2^{+}_{2}$ states etc. 
Combining OCM and GEM provides a powerful method to study the structure of light nuclei~\cite{yamada05,funaki08,yamada12} as well as light hypernuclei~\cite{hiyama97,hiyama09}, because of the following three points: 1)~the Pauli-blocking effect among the clusters is adequately taken into account by OCM, 2)~the GEM covers an approximately complete four-body model space~\cite{kamimura88,hiyama03}, and 3)~the cluster disintegrated thresholds of the $^{12}$C($0^+_1$,$2^+_1$,$4^+_1$,$0^+_2$,$3^-$,$1^-$)+$n$, $^{9}$Be($3/2^-_1$,$1/2^-_1$)+$\alpha$, $^{8}$Be($0^+$,$2^+$,$4^+$)+$^{5}$He($1/2^-$,$3/2^-$), and $3\alpha$+n channels etc.~are reasonably reproduced. 
The first point indicates that the OCM-GEM framework can describe the shell-model-like compact structure, for example, in the ground state of $^{13}$C.
For the third point, the reproduction of the cluster disintegrated thresholds is very important to discuss the cluster structures in $^{13}$C from the viewpoint of the Ikeda threshold rule~\cite{ikeda68}.
The present framework can explicitly treat a strong parity dependence of the $\alpha$-$n$ potential~\cite{kanada79}, demonstrated by the $\alpha$-$n$ scattering phase shifts~\cite{morgan68,brown67}:~the negative-parity potential is attractive enough to make resonant states ($3/2^{-}_{1}$, $1/2^{-}_{1}$, $7/2^{-}_{1}$ and $5/2^{-}_{1}$) of $^{5}$He, wheres the positive-parity potential is weakly repulsive. 
This strong parity dependence plays an important role in producing the cluster states of $^{13}$C, as will be discussed later. 
With this $3\alpha+n$ OCM framework, we investigate the cluster structures and monopole excitations as well as the Hoyle-analogue states in $^{13}$C.

The present paper is organized as follows:~In Sec.~\ref{sec:formulation}, the four-body $3\alpha+n$ OCM is formulated. Results and discussions are devoted to Sec.~\ref{sec:results_discussion}. Finally we present a summary in Sec.~\ref{sec:summary}.

\section{Formulation}\label{sec:formulation}

In this section, we present the formulation of the $3\alpha+n$ four-body OCM.

\begin{figure}[t]
\begin{center}
\includegraphics*[width=0.4\hsize]{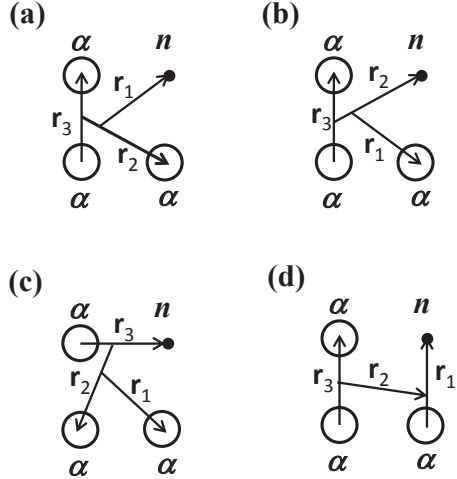}
\caption{
Jacobi-coordinate systems in the $3\alpha+n$ model.
(a)$\sim$(d) correspond to the respective ($p$-th) Jacobi coordinate systems ($p = 1 \sim 4$).
See the text.
}
\label{fig:1}
\end{center}
\end{figure} 
  
\subsection{$3\alpha+n$ OCM}\label{subsec:ocm}

The total wave function of $^{13}$C, $\tilde{\Psi}_J({^{13}{\rm C}})$, with the total angular momentum $J$ and total isospin $T=\frac{1}{2}$ in the  $3\alpha+n$ OCM framework is expressed by the product of the internal wave functions of $\alpha$ clusters $\phi(\alpha)$ and the relative wave function  $\Psi_J({^{13}{\rm C}})$ among the $3\alpha$ clusters and the extra neutron,
\begin{eqnarray}
\tilde{\Phi}_J({^{13}{\rm C}}) = \Phi_J({^{13}{\rm C}}) \phi^{\rm}(\alpha_1) \phi(\alpha_2) \phi(\alpha_3).
\end{eqnarray}
The relative wave function $\Phi_J({^{13}{\rm C}})$ is expanded in terms of the Gaussian basis as follows:
\begin{eqnarray}&&\Phi_J({^{13}{\rm C}})= \sum_{p=1}^{4} \sum_{c^{(p)}} \sum_{\nu^{(p)}}f^{(p)}_{c^{(p)}}(\nu^{(p)}) \Phi^{(p)}_{c^{(p)}}(\nu^{(p)}), 
\label{eq:total_wf} \\
&&\Phi^{(p)}_{c^{(p)}}(\nu^{(p)}) = \mathcal{S}_{\alpha} \left[ \left[ \varphi_{\ell^{(p)}_1}(\vc{r}^{(p)}_1,\nu^{(p)}_1) \left[ \varphi_{\ell^{(p)}_2}(\vc{r}^{(p)}_2,\nu^{(p)}_2) \varphi_{\ell^{(p)}_3}(\vc{r}^{(p)}_3,\nu^{(p)}_3) \right]_{L^{(p)}_{23}} \right]_{L^{(p)}} \xi_{\frac{1}{2}}(n) \right]_{J},
\label{eq:total_wf_basis}\\
&& {\langle u_F | \Phi_J({^{13}{\rm C}}) \rangle } = 0, \label{eq:oc}
\end{eqnarray}
where we assign the cluster number, 1, 2, and 3, to the three $\alpha$ clusters (spin 0), and the number 4 to the extra neutron (spin $\frac{1}{2}$). 
$\Phi^{(p)}_{c^{(p)}}(\nu^{(p)})$ denotes the relative wave function with respect to the $p$-th Jacobi-coordinate system of the four-body $3\alpha+n$ model (either of coordinate type of {\it K} or {\it H}) shown in Fig.~\ref{fig:1}, in which $\vc{r}^{(p)}_1$, $\vc{r}^{(p)}_2$, and $\vc{r}^{(p)}_3$ are the Jacobi coordinates. 
$\mathcal{S}_{\alpha}$ stands for the symmetrization operator acting on all $\alpha$ particles obeying Bose statistic, and $\xi_{\frac{1}{2}}(n)$ is the spin function of the extra neutron. 
$\phi^{\rm}(\alpha)$ denotes the intrinsic wave function of the $\alpha$ cluster with the $(0s)^{4}$ shell-model configuration. 
The angular momentum channel $c^{(p)}$ for the $p$-th Jacobi-coordinate system is presented as $c^{(p)}=\left[ \left[\ell^{(p)}_1, \left[\ell^{(p)}_2, \ell^{(p)}_3\right]_{L^{(p)}_{23}}\right]_{L^{(p)}}, \frac{1}{2}\right]_J$, where $\ell^{(p)}_i$ denotes the relative orbital angular momentum with respect to $\vc{r}^{(p)}_{i}$ in Fig.~\ref{fig:1} with $\vc{L}^{(p)}_{23}=\vc{\ell}^{(p)}_2+\vc{\ell}^{(p)}_3$, and $L^{(p)}$ stands for the total orbital angular momentum of the $3\alpha+n$ system with the total intrinsic spin $\frac{1}{2}$.
$\nu^{(p)}$ denotes the set of size parameters, $\nu^{(p)}_1$, $\nu^{(p)}_2$, and $\nu^{(p)}_3$, of the normalized Gaussian function, $\varphi_\ell(\vc{r},\nu)=N_\ell(\nu)r^\ell \exp(-\nu r^2) Y_\ell (\hat{\vc{r}})$, and $\nu$ is taken to be of geometrical progression,
\begin{equation}
\nu_n=1/b_n^2,\hspace{1cm}b_n=b_{\rm min}a^{n-1},\hspace{1cm}n=1\sim n_{\max}.
\label{eq:para_Gaussian} 
\end{equation}
It is noted that this prescription is found to be very useful in optimizing the ranges with a small number of free parameters ($b_{\rm min}$, $a$, $n_{\rm max}$) with high accuracy~\cite{kamimura88,hiyama03}.
Equation~(\ref{eq:oc}) represents the orthogonality condition that the total wave function (\ref{eq:total_wf}) should be orthogonal to the Pauli-forbidden states of the $3\alpha+n$ system, $u_F$'s, which are constructed from the Pauli forbidden states between two $\alpha$ particles in $0S$, $0D$, and $1S$ states together with those between $\alpha$ particle and extra neutron $n$ in $0S$ state~\cite{horiuchi77}.
In the present study, the Pauli-forbidden states are removed with use of the Kukulin's method~\cite{kukulin} as shown later. 
The ground state with the dominant shell-model-like configuration $(0s)^{4}(0p)^{9}$, then, can be properly described in the present $3\alpha+n$ OCM framework.

The $3\alpha+n$ Hamiltonian for $\Phi_J({^{13}{\rm C}})$ is presented as
\begin{eqnarray}
\mathcal{H}&=&\sum_{i=1}^{4} T_i-T_{\rm cm} + \sum_{i<j=1}^{3} V_{{2\alpha}}(i,j) + \sum_{i=1}^{3} V_{{\alpha}n}(i,4)
\nonumber \\
           &+&  V_{3\alpha}(1,2,3) + \sum_{i<j=1}^{3} V_{2\alpha n}(i,j,4) + V_{3\alpha n}(1,2,3,4) + V_{\rm Pauli},
\label{eq:hamiltonian}
\end{eqnarray}
where $T_i$, $V_{2\alpha}$ ($V_{\alpha n}$), $V_{3\alpha}$ ($V_{2\alpha n}$), and $V_{3\alpha n}$ stand for the kinetic energy operator for the {\it i}-th cluster, $\alpha$-$\alpha$ ($\alpha$-$n$) potential, three-body potential among the three $\alpha$ particles (the two $\alpha$ particles and extra neutron), and four-body potential, respectively. The center-of-mass kinetic energy ($T_{\rm cm}$) is subtracted from the Hamiltonian. 

\begin{figure}[t]
\begin{center}
\includegraphics[clip,width=0.90\hsize]{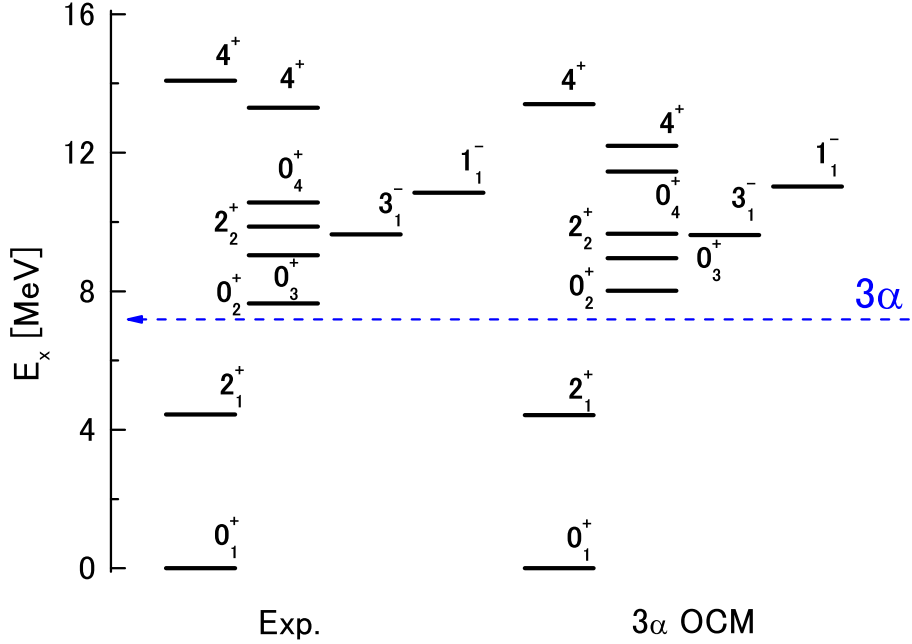}
\caption{
(Color online) Energy spectra of $^{12}$C obtained by the $3\alpha$ OCM calculation with the three-body force $V_{3\alpha}$ in Eq.~(\ref{eq:3a_force}), compared with the experimental data (see the text for the details).
}
\label{fig:2}
\end{center}
\end{figure}

The effective $\alpha$-$\alpha$ potential $V_{2\alpha}$ is constructed by the folding procedure from an effective two-nucleon force including the proton-proton Coulomb force. Here we take the Schmid-Wildermuth (SW) force~\cite{sw} as the effective $NN$ force. 
It is noted that the folded $\alpha$-$\alpha$ potential reproduces the $\alpha$-$\alpha$ scattering phase shifts and the energies of the $^8$Be ground-band state ($J^{\pi}=0^{+}-2^{+}-4^{+}$) within the framework of the $2\alpha$ OCM. 
As mentioned in Sec.~\ref{sec:introduction}, the $3\alpha$ OCM calculations with the complex-scaling method by Kurokawa et al.~\cite{kurokawa05} and Ohtsubo et al.~\cite{ohtsubo13}, who use the $\alpha$-$\alpha$ folding potential based on the SW force together with the phenomenological effective three-body force $V_{3\alpha}$ depending on the total angular momentum of $^{12}$C, reproduce the $2^{+}_2$ and $4^{+}_2$ states, which have been recently identified above the Hoyle state~\cite{itoh04,freer09,itoh11,fynbo11,zimmerman11,zimmerman13}, and the $0^{+}_3$ and $0^{+}_4$ states, which have quite recently been observed in experiments~\cite{itoh11}.
The origin of $V_{3\alpha}$ is thought to be due to the state dependence of the effective nucleon-nucleon interaction and an additional Pauli repulsion arising from exchanging nucleons among the $3\alpha$ clusters.   
In order to incorporate the angular momentum dependence of $V_{3\alpha}$ given in Refs.~\cite{kurokawa05,ohtsubo13} into the present $3\alpha+n$ OCM framework as possible, we take the following simple one-range Gaussian  
\begin{eqnarray}
V_{3\alpha}(1,2,3) =   V_{3\alpha}^{(0)} \exp\left[-{\beta_{3\alpha}} (\vc{\rho}^2_{12}+\vc{\rho}^2_{23}+\vc{\rho}^2_{31}) \right] \left( 1  + \eta_{3\alpha} \vc{L}^2_{3\alpha} \right)
\label{eq:3a_force}
\end{eqnarray}
where the operator $\vc{L}_{3\alpha}$ stands for the total orbital angular momentum operator of the $3\alpha$ clusters in the $3\alpha+n$ system, and $\vc{\rho}_{ab}$ denotes the relative coordinate between the $a$-th $\alpha$ particle and $b$-th one.
In the present study we use $\beta_{3\alpha}=0.15$~fm$^{-2}$, $V_{3\alpha}^{(0)}=31.7$~MeV, and $\eta_{3\alpha}=(63.0/31.7 -1)/6$, where the value of $\beta_{3\alpha}$ is the same as that used in Refs.~\cite{kurokawa05,ohtsubo13}.
These parameter sets give the strength of $V_{3\alpha}^{(0)}\left( 1  + \eta_{3\alpha} \vc{L}^2_{3\alpha} \right)=31.7,~63.0$, and $136.0$~MeV for the total angular momentum $L_{3\alpha}=0,2$, and $4$, respectively, which are comparable to the strengths of $V_{3\alpha}=31.7,~63.0$, and $150$~MeV used in Refs.~~\cite{kurokawa05,ohtsubo13}.
The calculated energy spectra of $^{12}$C with the $3\alpha$ OCM using the folding potential based on the SW force and three-body force $V_{3\alpha}$ in Eq.~(\ref{eq:3a_force}) are shown in Fig.~\ref{fig:2}.
The resonant states above the $3\alpha$ threshold are obtained by applying the complex scaling method (CSM) to the $3\alpha$ OCM, although only the $0^+_3$ resonant state is identified with the more sophisticated method, i.e.,  the method of analytic continuation in the coupling
constant combined with the complex scaling method (ACCC+CSM) proposed by Ref.~\cite{kurokawa05}.
Compared with the experimental data, they are reasonably reproduced.
It is known that the $3\alpha$ OCM describes well the structure of the $0^{+}_{1,2}$, $2^{+}_{1,2}$, $4^{+}_{1,2}$, $3^-$, and $1^-$ states in $^{12}$C.

\begin{figure}[t]
\begin{center}
\includegraphics[clip,width=0.8\hsize]{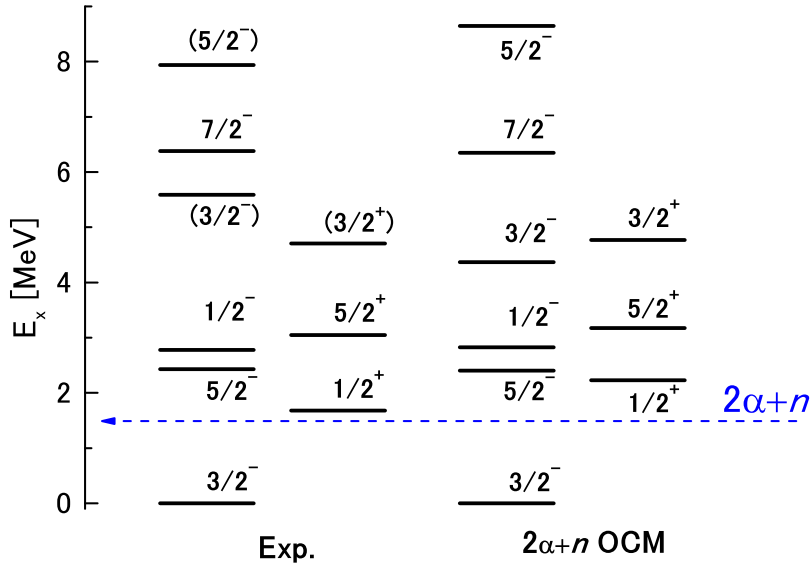}
\caption{
(Color online) Low-lying energy levels of $^9$Be obtained by the $2\alpha+n$ OCM calculation together with the experimental data (see the text for the details).
}
\label{fig:3}
\end{center}
\end{figure}

As for the effective $\alpha$-$n$ potential $V_{\alpha n}$ in Eq.~(\ref{eq:hamiltonian}), we use the Kanada-Kaneko potential~\cite{kanada79} which reproduces nicely the $\alpha$-$n$ scattering phase shifts with the odd and even partial waves in the low-energy region together with the resonant states ($J^{\pi}=3/2^{-}_1$ and $1/2^-_1$) of $^{5}$He. 
This local potential, which has the parity-dependent term, is constructed within the microscopic $\alpha$+$n$ cluster model based on the resonating group method (RGM). 
In the Hamiltonian (\ref{eq:hamiltonian}), the three-body force $V_{2\alpha{n}}$ of the short-range Gaussian type is phenomenologically introduced so as to fit the ground-state energy of $^9$Be measured from the $2\alpha+n$ threshold within the framework of the $2\alpha+n$ OCM. 
The origin of $V_{2\alpha{n}}$ is considered to be the same reason as the case of $V_{3\alpha}$ mentioned above.
In the present paper, we take the following phenomenological Gaussian-type potential,
\begin{eqnarray}
&&V_{2\alpha{n}}(i,j,4) =  V_{2\alpha{n}}^{(0)} \exp\left[-{\beta^2_{2\alpha{n}}} (\vc{\rho}^2_{ij}+\vc{\rho}^2_{i4}+\vc{\rho}^2_{j4}) \right],
\end{eqnarray}
where $\vc{\rho}_{ab}$ denotes the relative coordinate between $a$-th and $b$-th particles, and $(i,j)$ is either $(1,2)$, $(1,3)$, or $(2,3)$.
For simplicity we take $\beta_{2\alpha{n}}=\beta_{3\alpha}=0.15$~fm$^{-2}$.
The calculated energy spectra of $^9$Be with the $2\alpha+n$ OCM are shown in Fig.~\ref{fig:3}. 
The resonant states above the $2\alpha+n$ threshold are obtained by applying the CSM to the $2\alpha+n$ OCM, although only the $1/2^+$ state just above the $2\alpha+n$ threshold is given with the bound state approximation.
It is noted that the $1/2^+$ state is considered to be a virtual state~\cite{okabe77} and thus the CSM fails to fix its identification as shown in Ref.~\cite{arai03}.
Compared with the experimental data, the low-lying spectra in $^{9}$Be are reasonably reproduced. 
It is known that the $2\alpha+n$ OCM describes well the cluster structure of $^{9}$Be in the low-lying region.

As for the four-body force $V_{3\alpha n}$ in Eq.~(\ref{eq:hamiltonian}), a phenomenological one is introduced so as to reproduce the energies of the ground state ($1/2^-_1$) with respect to the $3\alpha+n$ threshold. 
The origin of $V_{3\alpha n}$ is similar to that in the case of the three-body forces  $V_{3\alpha}$ and $V_{2\alpha n}$ as mentioned above. 
It should be short-range, and hence only act in compact configurations. 
In the present paper, we take the following phenomenological four-body potential,
\begin{eqnarray}
V_{3\alpha{\it n}} =  \sum_{Q=9} V_0(Q) \sum_{(\lambda,\mu)} \sum_{L^{\pi}}
  {|\Phi^{\rm SU(3)}_{(\lambda,\mu)Q}({L^{\pi}})\rangle} {\langle\Phi^{\rm SU(3)}_{(\lambda,\mu)Q}(L^{\pi})|},
\end{eqnarray}
where $\Phi^{\rm SU(3)}_{(\lambda,\mu)Q}({L^{\pi}})$ with the total orbital angular momentum $L$ represents the SU(3)$[4441](\lambda,\mu)$ wave function with the total harmonic oscillator quanta $Q$ ($Q=9$). 
It is noted that the present $3\alpha+n$ model space can be classified into the SU(3) bases with the irreducible representation, $(\lambda,\mu)Q$, with partition $[f]=[4441]$, and the total wave function of $^{13}$C with negative (positive) parity in Eq.~(\ref{eq:total_wf}) can be expanded in terms of the $\Phi^{\rm SU(3)}_{(\lambda,\mu)Q}({L^{\pi}})$ bases with odd (even) $Q$ values. 
The $(\lambda,\mu)=(0,3)$ basis with $Q=9$ is the unique Pauli allowed state for the negative-parity state of $^{13}$C with $Q=9$, which is equivalent to the shell-model configuration of $(0s)^4(0p)^{9}$. 
Thus this SU(3) basis turns out to be the main component in the ground state of $^{13}$C. 
For simplicity, the strength of the four-body potential, $V_0(Q=9)$, is fixed so as to reproduce the experimental energy of the ground state of $^{13}$C with respect to the $3\alpha+n$ threshold:~$V_0(Q=9)=2.0$~MeV. 
The expectation value of this four-body potential does not exceed $3~\%$ of that of the corresponding two-body and three-body terms, even for the ground state with the most compact structure, i.e.~being the most sensitive to the potential. 
This type of potential based on the SU(3) framework is also used as the three-body potential in the $\alpha+\alpha+t$ OCM calculation of $^{11}$B~\cite{yamada10}.

The Pauli-blocking operator $V_{\rm Pauli}$ in Eq.~(\ref{eq:hamiltonian}), which is based on the Kukulin's method~\cite{kukulin}, is expressed as 
\begin{eqnarray}
V_{\rm Pauli} = \lim_{\lambda \rightarrow  \infty}\ {\lambda}\ \sum_{f}\ {| u_f \rangle}{\langle u_f |},
\end{eqnarray} 
which rules out the Pauli-forbidden $\alpha$-$\alpha$ relative states ($f=0S,1S,0D$) and Pauli-forbidden $\alpha-n$ relative state ($f=0S$) from the four-body $3\alpha-n$ wave function.
In the present study, we take $\lambda=10^{4}$~MeV.

The equation of motion of $^{13}$C with the $3\alpha+n$ OCM is obtained by the variational principle,
\begin{eqnarray}
\delta\left[\langle\Phi_J({^{13}{\rm C}})\mid \mathcal{H}-E \mid \Phi_J({^{13}{\rm C}})\rangle\right]=0,
\label{eq:variational_principle}
\end{eqnarray}
where $E$ denotes the eigenenergy of $^{13}$C measured from the $3\alpha+n$ threshold. The energy $E$ and expansion coefficients $f^{(p)}_{c^{(p)}}$  in the total wave function shown in Eq.~(\ref{eq:total_wf}) are determined by solving a secular equation derived from Eq.~(\ref{eq:variational_principle}). 

It is instructive to study single-$\alpha$-particle orbits and corresponding occupation probabilities in $^{13}$C. 
We define the single-cluster density matrix for $\alpha$ clusters, respectively, as
\begin{eqnarray}
&&\rho^{(\alpha)}(\vc{r},\vc{r}')
 = \langle \Phi_J({^{13}{\rm C}}) |~\frac{1}{3}\sum_{i=1}^{3} {|\delta(\vc{r}^{(G)}_i-\vc{r}')\rangle}{\langle\delta(\vc{r}^{(G)}_i-\vc{r})|}~|\Phi_J({^{13}{\rm C}})\rangle,
\label{eq:single_alpha_density}
\end{eqnarray}
where $\vc{r}^{(G)}_i$ $(i=1,2,3)$  represents the coordinate vector of the $i$th $\alpha$ cluster with respect to the center-of-mass coordinate of the $3\alpha+n$ system. 
The calculated method of $\rho$ is given in Refs.~\cite{yamada05,matsumura04,suzuki02,yamada08_density}. 
The single-$\alpha$-particle orbits and corresponding occupation probabilities are obtained by solving the eigenvalue equation of the single-cluster density matrix,
\begin{eqnarray}
&&\int d\vc{r} \rho^{(\alpha)}(\vc{r},\vc{r}') f^{(\alpha)}_\mu(\vc{r}') = \mu^{(\alpha)} f^{(\alpha)}_\mu(\vc{r}),
\label{eq:density_alpha}
\end{eqnarray}
where the eigenvalue $\mu^{(\alpha)}$ denotes the occupation probability for the corresponding single-cluster orbit $f^{(\alpha)}_\mu$ with the argument of the intrinsic coordinate of an arbitrary $\alpha$ cluster  measured from the center-of-mass coordinate of $^{13}$C. 
The spectrum of the occupation probabilities provides important information on the occupancies of the single-$\alpha$-particle orbit in $^{13}$C. 
If the three $\alpha$ particles occupy only an single orbit, the occupation probability for this orbit becomes 100~\%.
On the other hand, the single-particle density matrix for the extra neutron in $^{13}$C is also defined as 
\begin{eqnarray}
&&\rho^{(n)}(\vc{r},\vc{r}')
 = \langle \Phi_J({^{13}{\rm C}}) |~{|\delta(\vc{r}^{(G)}_4-\vc{r}')\rangle}{\langle\delta(\vc{r}^{(G)}_4-\vc{r})|}~|\Phi_J({^{13}{\rm C}})\rangle,
\label{eq:single_nucleon_density}
\end{eqnarray}
where $\vc{r}^{(G)}_4$  stands for the coordinate vector of the extra neutron with respect to the center-of-mass coordinate of the $3\alpha+n$ system. 
The single-particle orbits and corresponding occupation probabilities of the extra neutron are obtained by diagonalizing the density matrix in the same manner as the case of the single-$\alpha$-cluster density matrix (see Eqs.~(\ref{eq:single_alpha_density}) and (\ref{eq:density_alpha})). 

The root-mean-square (rms) radius of $^{13}$C in the present OCM is given as
\begin{eqnarray}
R &=& \left\langle \frac{1}{13}\sum_{i=1}^{13} (\vc{r}_i - \vc{R}_{\rm cm})^2 \right\rangle^{1/2} \nonumber \\
       &=& \left[ \frac{1}{13}\left( 12{\langle r^2 \rangle}_{\alpha} +  2{R^2_{\alpha-\alpha}} + \frac{8}{3} {R^2_{\alpha-{^8{\rm Be}}}} + \frac{12}{13}{R^2_{n-{^{12}{\rm C}}}} \right) \right]^{1/2},
\label{eq:rms}
\end{eqnarray}
where $R_{\alpha-\alpha}$ ($R_{\alpha-{^8{\rm Be}}}$, $R_{n-{^{12}{\rm C}}}$) presents the rms distance between $\alpha$ and $\alpha$ (the third $\alpha$ and $^8{\rm Be}$, extra neutron and $^{12}$C) in $^{13}$C. 
In Eq.~(\ref{eq:rms}) we take into account the finite size effect of $\alpha$ clusters, where the experimental rms radius for the $\alpha$ particle is used in $\sqrt{\langle r^2 \rangle}_{\alpha}$.

The overlap amplitudes or reduced width amplitude is useful to see the degree of clustering in nucleus. 
In the present paper, we study the reduced width amplitudes for the $^{12}$C+$n$ and $^9$Be+$\alpha$ channels, respectively, defined as
\begin{eqnarray}
&&{\cal Y}_{J_{C}(\ell_{n}\frac{1}{2})j_{n}J}^{{^{12}}{\rm C}-n}({r}_{n})
 =  \left\langle \left[ \frac{\delta({r^{\prime}_{n}}-r_{n})}{{r^{\prime}_{n}}^2} [Y_{\ell_{n}}(\hat{\vc{r}}^{\prime}_{n})\chi_{\frac{1}{2}}(n)]_{j_n} \phi_{J_{C}}{({^{12}}{\rm C})}  \right]_J | \tilde{\Phi}_J({^{13}{\rm C}}) \right\rangle,
\label{eq:RWA_neutron}\\
&&{\cal Y}_{J_9\ell_{94}J}^{{^9}{\rm Be}-\alpha}({r}_{94})
 = \sqrt{\frac{3!}{2!1!}} \left\langle \left[ \frac{\delta({r^{\prime}_{94}}-r_{94})}{{r^{\prime}_{94}}^2} Y_{\ell_{94}}(\hat{\vc{r}}^{\prime}_{94}) \phi_{J_9}({^9}{\rm Be}) \phi(\alpha)  \right]_J | \tilde{\Phi}_J({^{13}{\rm C}}) \right\rangle,
\label{eq:RWA_alpha}
\end{eqnarray}
where $r_{n}$ ($r_{94}$) denotes the radial part of the relative coordinate between $^{12}$C and $n$ ($^9$Be and $\alpha$). 
The wave function $\phi_{J_{C}}{({^{12}}{\rm C})}$ ($\phi_{J_9}({^9}{\rm Be})$) of $^{12}$C ($^{9}$Be) with the total angular momentum $J_C$ ($J_9$) is obtained with the $3\alpha$ OCM ($2\alpha+n$ OCM). 
The spectroscopic factor $S^2$ is defined as 
\begin{eqnarray}
S^{2}=\int_{0}^{\infty} dr [r \times {\cal Y}(r)]^2,
\label{eq:s2-factor}
\end{eqnarray}
where ${\cal Y}$ denotes the overlap amplitude.

\subsection{Monopole transitions, E1 transitions, and Gamow-Teller transitions}\label{subsec:is_e1_monopole_GT}

The {C0} (longitudinal electric monopole) transition matrix element between the ground state ($1/2^{-}_1$) and $n$-th excited $1/2^-$ state by the $(e,e\rq{})$ reaction is given as follows:
\begin{eqnarray}
{M({\rm C0},{1/2^{-}_{n}}-{1/2^{-}_{1}})} &=& {\langle  {\tilde{\Phi}_{1/2^{-}_{n}}({^{13}{\rm C}})} | \sum_{i=1}^{13} \frac{1+\tau_{3i}}{2}  (\vc{r}_i - \vc{R}_{\rm cm} )^2 | {\tilde{\Phi}_{1/2^{-}_{1}}({^{13}{\rm C}})} \rangle }, \label{eq:me_e0}
\end{eqnarray}
where $\vc{r}_i$ ($i=1\sim13$) are the coordinates of nucleons and $\vc{R}_{\rm cm}={\frac{1}{13}}\sum_{i=1}^{13}\vc{r}_{i}$ is the c.o.m.~coordinate of $^{13}$C. 
This is the same definition as the {C0} transition matrix element, which is related with the $e^+e^-$ pair creation process, for example, the 6.05 MeV $0^+_2 \rightarrow 0^+_1$ transition in $^{16}$O.
On the other hand, the isoscalar monopole transition matrix element from the ground state ($1/2_1^{-}$) to the $n$-th excited state ($1/2_{n}^{-}$) is defined as 
\begin{eqnarray}
\mathcal{M}({\rm IS},{1/2^{-}_{n}}-{1/2^{-}_{1}}) = {\langle {\tilde{\Phi}_{1/2^{-}_{n}}({^{13}{\rm C}})} | \sum_{i=1}^{13} (\vc{r}_i - \vc{R}_{\rm cm} )^2 | {\tilde{\Phi}_{1/2^{-}_{1}}({^{13}{\rm C}})} \rangle}.\label{eq:me_is_monopole}
\end{eqnarray}
In the present study we calculate the ${\rm E1}$ transition rate between the ground state ($1/2^-_1$) and first $1/2^+$ state ($1/2^+_1$) together with the Gamow-Teller transition rate between the ground state of $^{13}$C and $1/2^{-}_1$ ($3/2^-_1$) state of $^{13}$N, 
\begin{eqnarray}
&&B({\rm E1})=|\langle \tilde{\Phi}_{1/2^{+}_{1}}({^{13}{\rm C}}) || \sum_{i=1}^{13} \frac{1+\tau_{3i}}{2} |\vc{r}_i - \vc{R}_{\rm cm}| Y_1(\widehat{\vc{r}_i - \vc{R}_{\rm cm}}) || \tilde{\Phi}_{1/2^{-}_{1}}({^{13}{\rm C}}) \rangle|^2, 
\label{eq:E1}\\
&&B({\rm GT})=|\langle \tilde{\Phi}_{J_f=1/2^{-}_1,3/2^{-}_1}({^{13}{\rm N}}) || \sum_{i=1}^{13}\sigma_i\tau_i || \tilde{\Phi}_{1/2^{-}_{1}}({^{13}{\rm C}}) \rangle|^2. 
\label{eq:GT}
\end{eqnarray}
The wave functions of the ground state ($1/2^-$) and first $3/2^-$ states of $^{13}$N, $\tilde{\Phi}_{J_f=1/2^{-}_1,3/2^{-}_1}({^{13}{\rm N}})$, are obtained by the four-body $3\alpha+p$ OCM calculation, in which only the Coulomb forces between the extra proton and three $\alpha$ clusters are switched on in the present formulation of the $3\alpha+n$ OCM (see Sec.~\ref{subsec:ocm}), although the details will be given elsewhere.

\section{Results and discussion}\label{sec:results_discussion}

\begin{figure}[t]
\begin{center}
\includegraphics*[width=0.88\hsize]{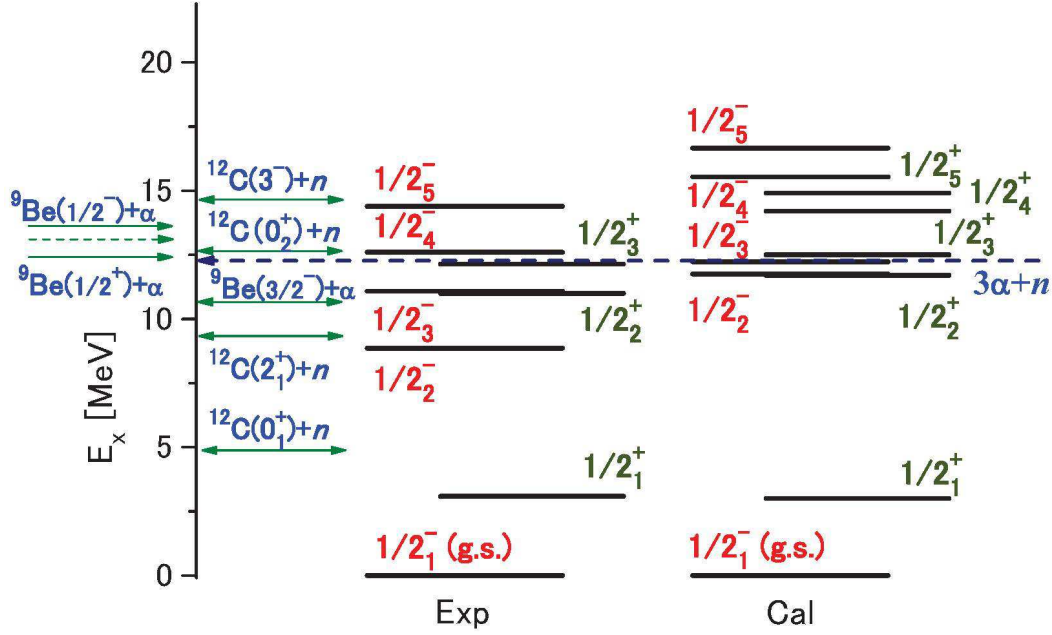}
\caption{
(Color online) Energy levels of the $1/2^-$ and $1/2^+$ states of $^{13}$C obtained by the $3\alpha+n$ OCM, compared with the experimental data.
The experimental levels of the $1/2^{-}_{1,2,3,5}$ and $1/2^+_{1,2,3}$ states are taken from Ref.~\cite{ajzenberg93} and from Ref.~\cite{kawabata08} for the $1/2^{-}_{4}$ state, respectively.
The threshold of the $^9$Be($5/2^{-}$)+$\alpha$ channel at $E_x=13.1$~MeV, located between the  $^9$Be($1/2^{+}$)+$\alpha$ and $^9$Be($1/2^{-}$)+$\alpha$ channels, is presented by the dashed arrow in the left hand side of the panel. 
}
\label{fig:4}
\end{center}
\end{figure}

Figure~\ref{fig:4} shows the energy levels of $1/2^{-}$ states together with those of $1/2^+$ states in $^{13}$C obtained by the four-body $3\alpha+n$ OCM calculation.
We found that five $1/2^-$ states and five $1/2^+$ states come out as either bound states against particle decays or quasi-bound states.

\subsection{$1/2^{-}$ states}\label{subsec:structures_12minus}

First we discuss the structures of the five $1/2^{-}$ states together their isoscalar monopole excitations and ${\rm C0}$ transitions.

\subsubsection{Structures of $1/2^{-}$ states}\label{subsub:structures_12minus}
 
The $1/2^{-}_1$ state, located at $E=-12.3$ MeV measured from the $3\alpha+n$ threshold, is the ground state of $^{13}$C.
Its calculated r.m.s.~radius is $R_N=2.4$ fm (see Table~\ref{tab:12minus_ocm}), the value of which is in correspondence with the experimental data ($2.46$ fm).
According to the analysis of the wave function, the main component of this state is the SU(3) irreducible representation $[f](\lambda,\mu)=[4441](0,3)$ of the $(0s)^{4}(0p)^{9}$ configuration with the lowest harmonic oscillator quanta $Q=9$ and its dominant angular momentum channel is $(L,S)_J=(1,\frac{1}{2})_{\frac{1}{2}}$, where $L$ ($S$) denotes the total orbital angular momentum (total intrinsic spin):~
\textcolor{black}{
The component of the SU(3) $(0,3)$ state ($0\hbar\omega$) is as large as $61~\%$, and the remaining comes from the $2\hbar\omega$ state ($18~\%$), $4\hbar\omega$ ($11~\%$), $6\hbar\omega$ ($5~\%$), and higher $\hbar\omega$ ($5~\%$).
In the cluster model, the components other than the $0\hbar\omega$ one correspond to $\alpha$-type ground-state correlations~\cite{yamada08_monopole,yamada12}.
}
It is noted that the SU(3) state, $[4441](0,3)$, is the lowest good-spatial-symmetry state in the SU(3) model of $^{13}$C. 
Reflecting the fact that the nuclear force favors a good spatial symmetry, the ground state of $^{13}$C has the dominant SU(3)-like nature,
\textcolor{black}{
although it has significant $\alpha$-type ground-state correlations in the present cluster model.
}

\begin{figure}[t]
\begin{center}
\includegraphics[width=50mm]{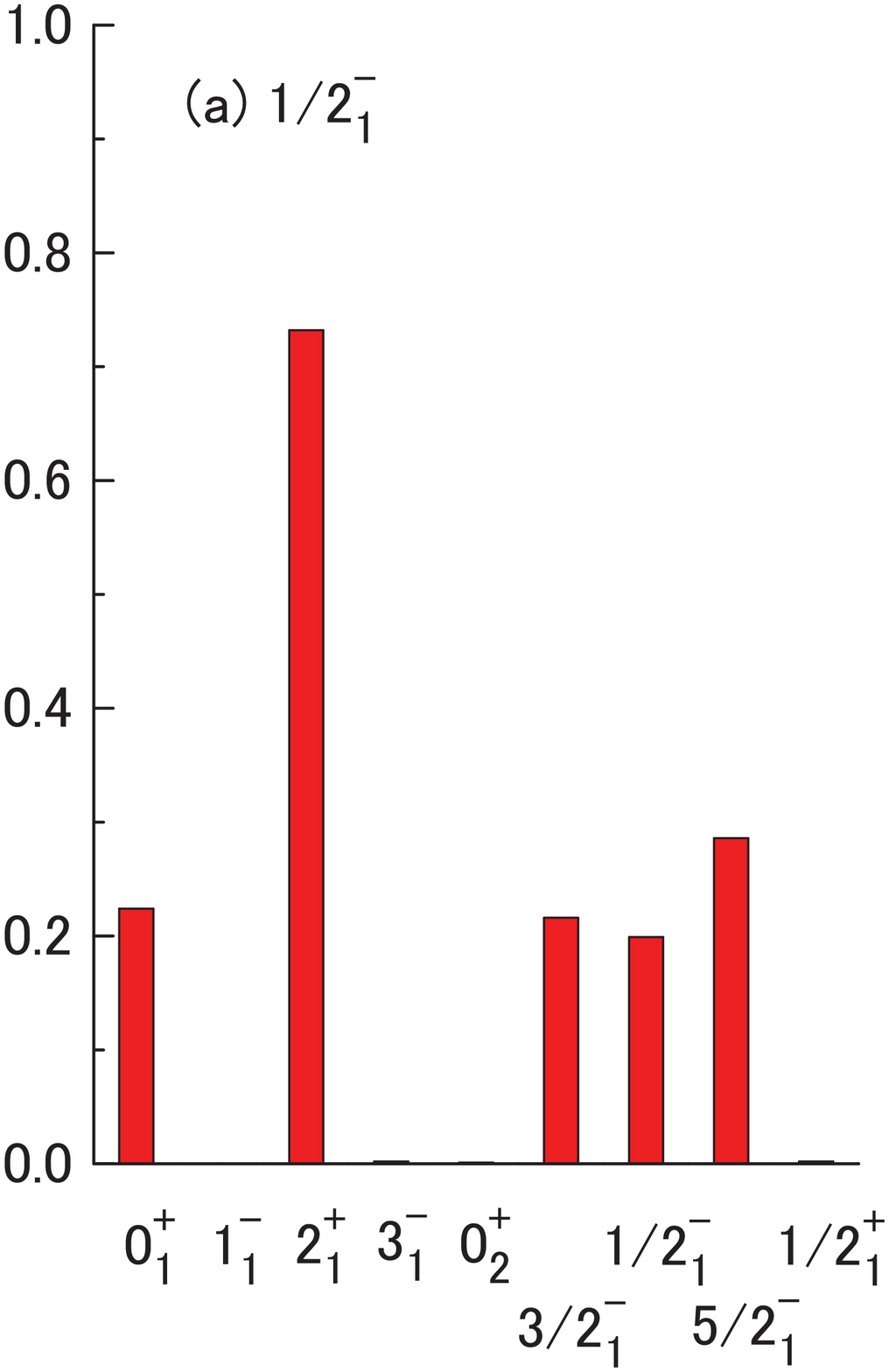}
\includegraphics[width=50mm]{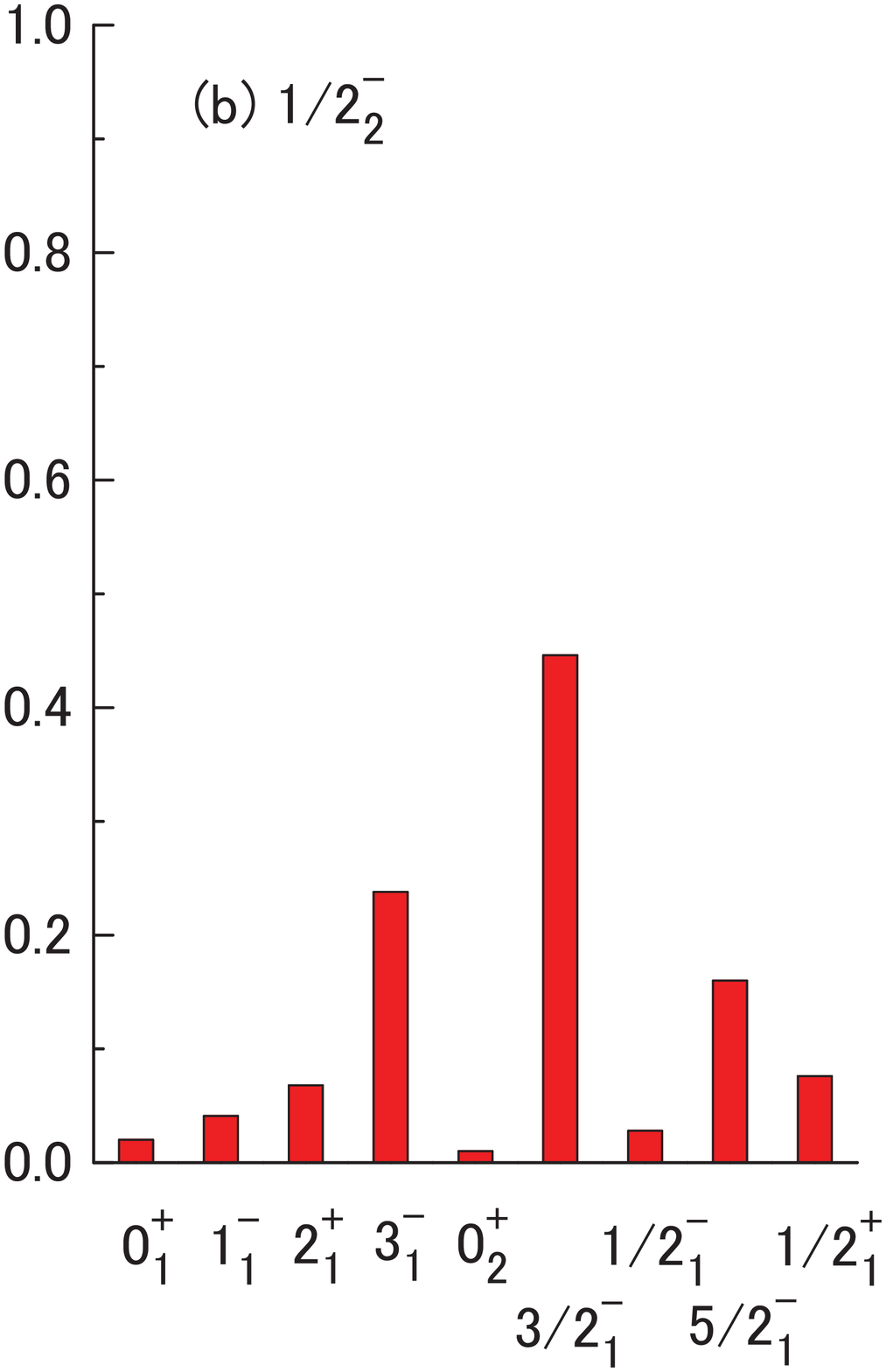}
\includegraphics[width=50mm]{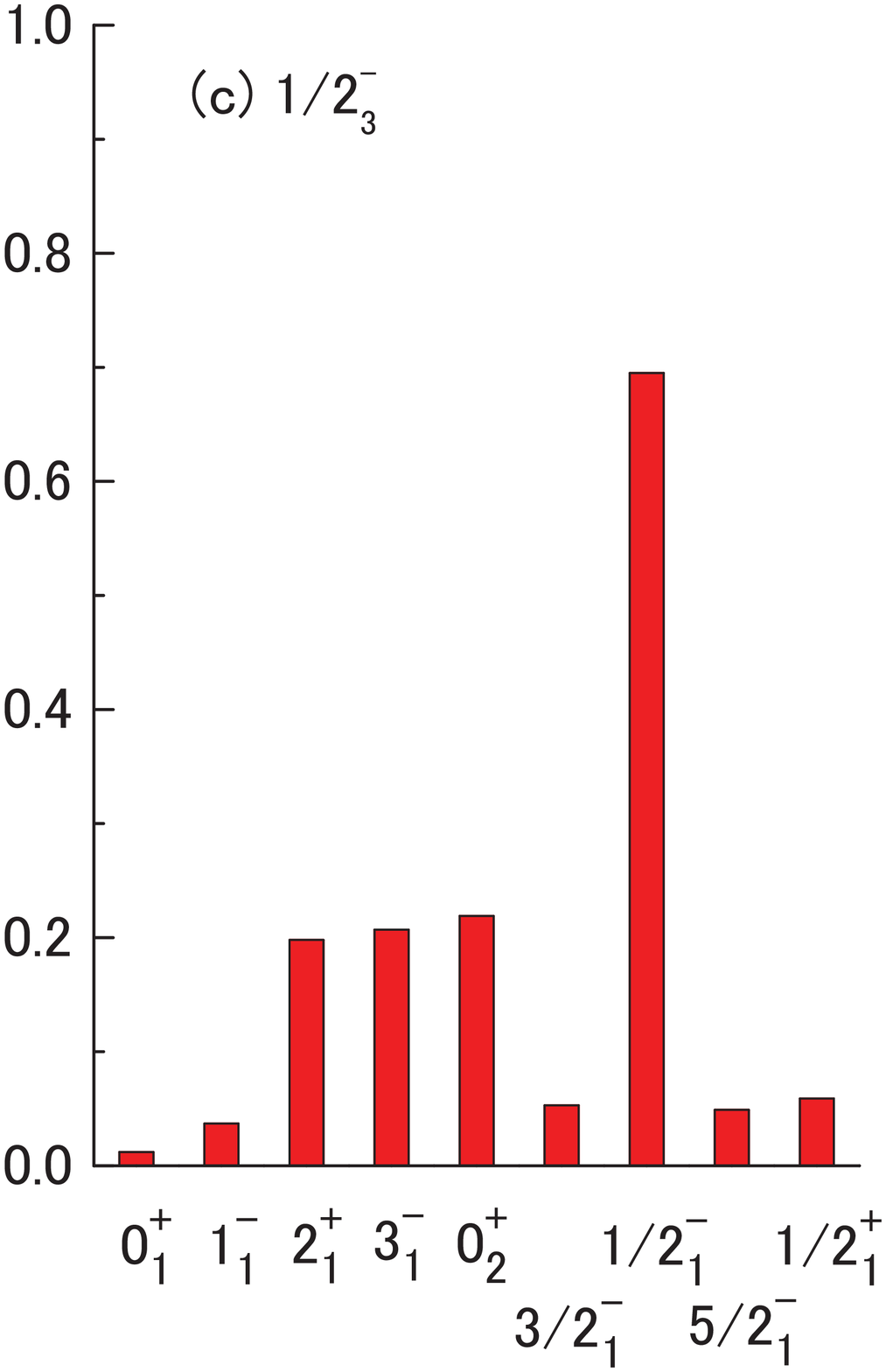}
\\ \vspace*{5mm}
\includegraphics[width=50mm]{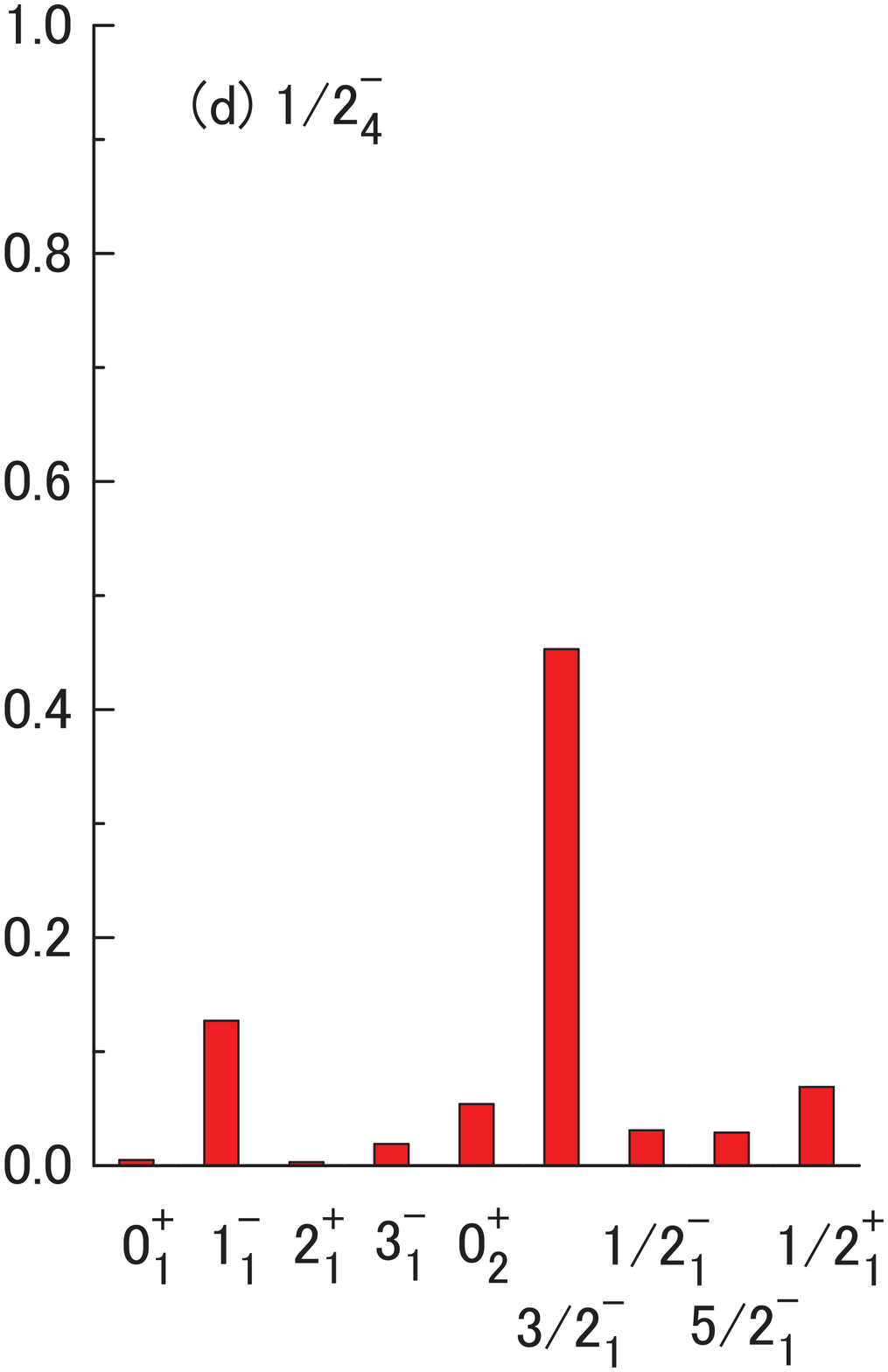}
\includegraphics[width=50mm]{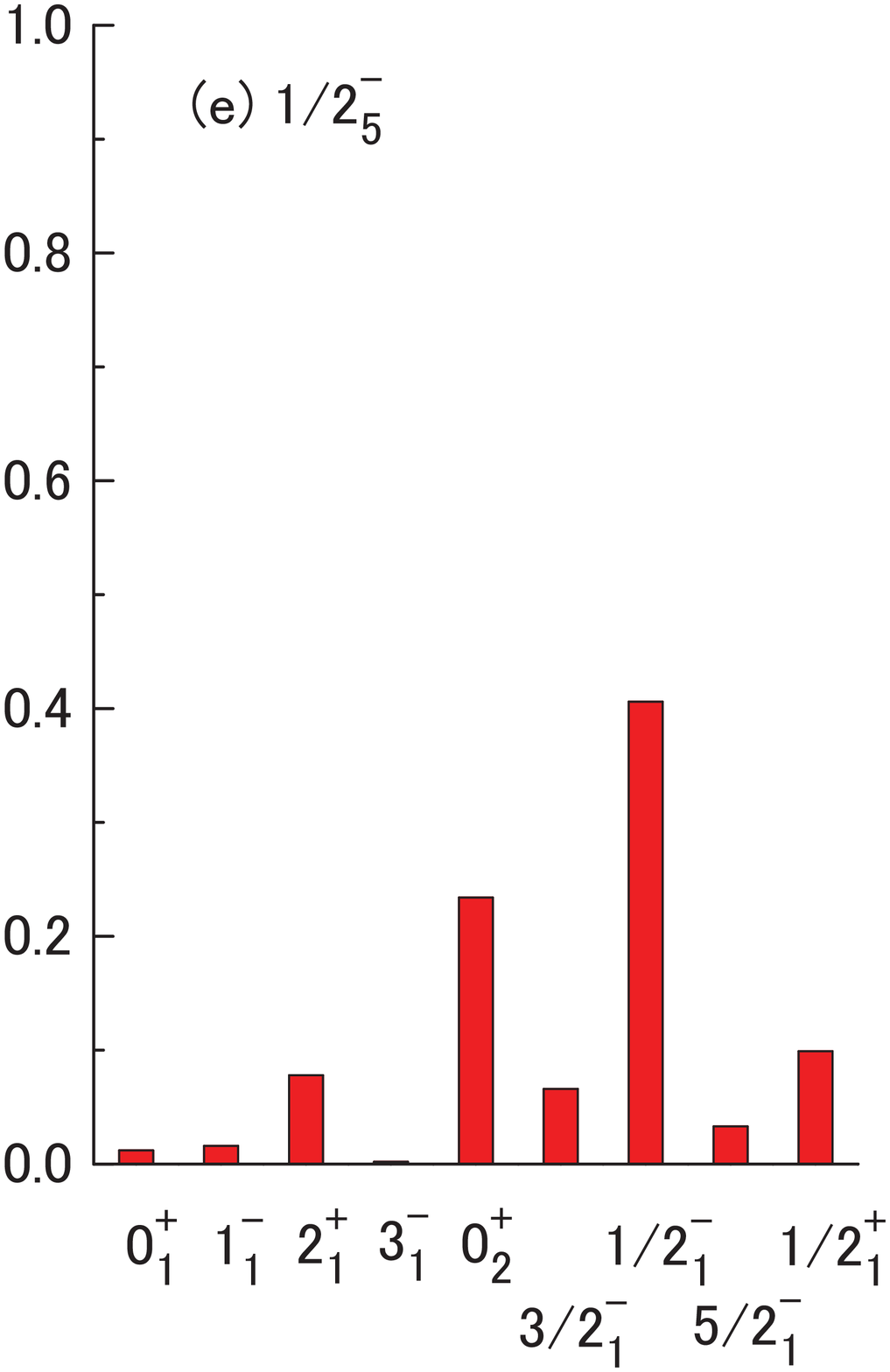}
\caption{(Color online) {Spectroscopic factors $S^{2}_{L}$} of the $^{12}$C$(J^{\pi}_C)$+$n$ channels ($J^{\pi}_C=0^{+}_{1},1^{-}_{1},2^{+}_{1},3^{-}_{1},0^{+}_{2}$) and $^{9}$Be($J^{\pi}_9$)+$\alpha$ ($J^{\pi}_9=3/2^{-}_1,1/2^{-}_1,5/2^{-}_1,1/2^{+}_1$) in the five $1/2^{-}$ states of $^{13}$C {defined in Eq.~(\ref{eq:s2-factor})}.}
\label{fig:s2_factors_12_minus_13c}
\end{center}
\end{figure}

\begin{figure}[t]
\begin{center}
\includegraphics[width=81mm]{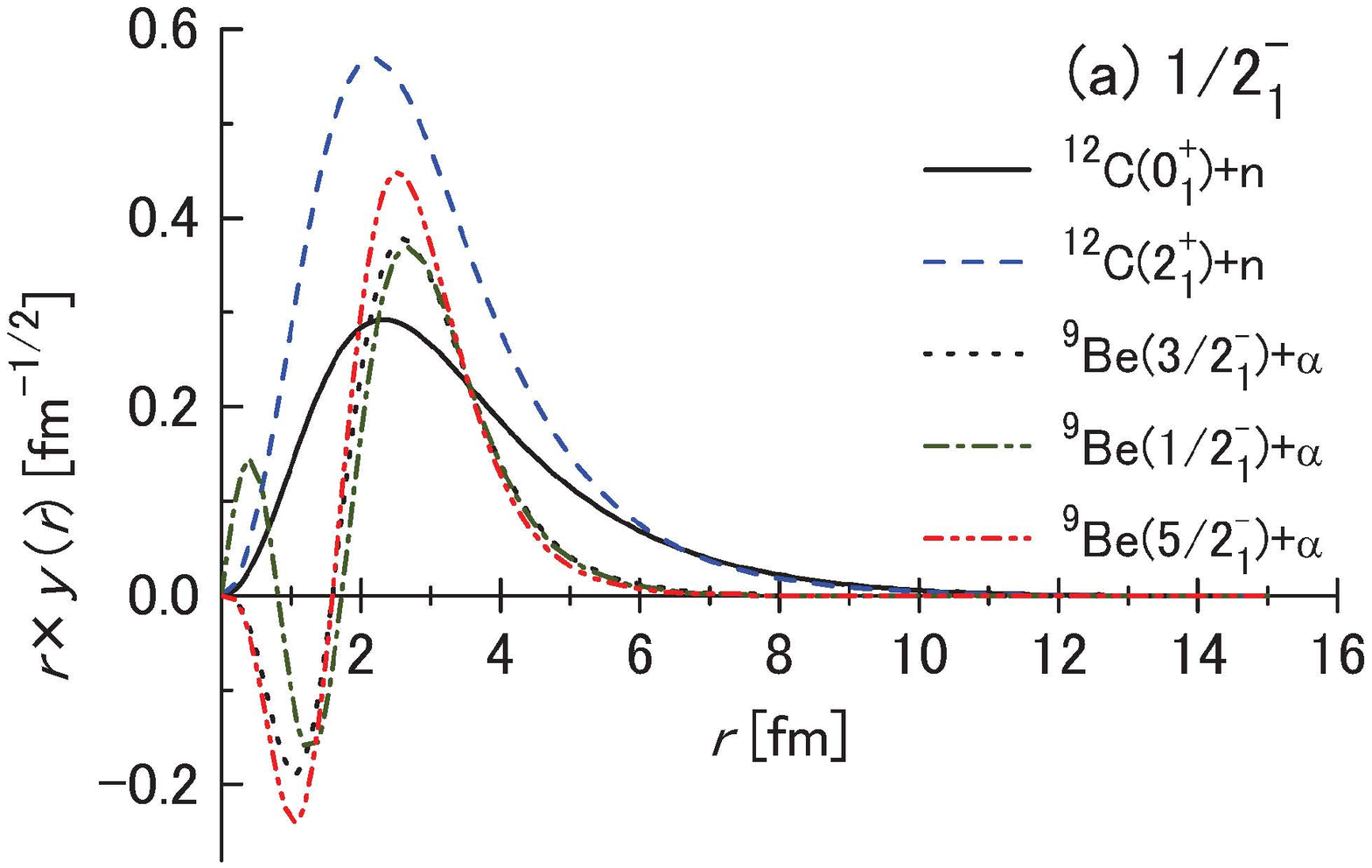}
\includegraphics[width=81mm]{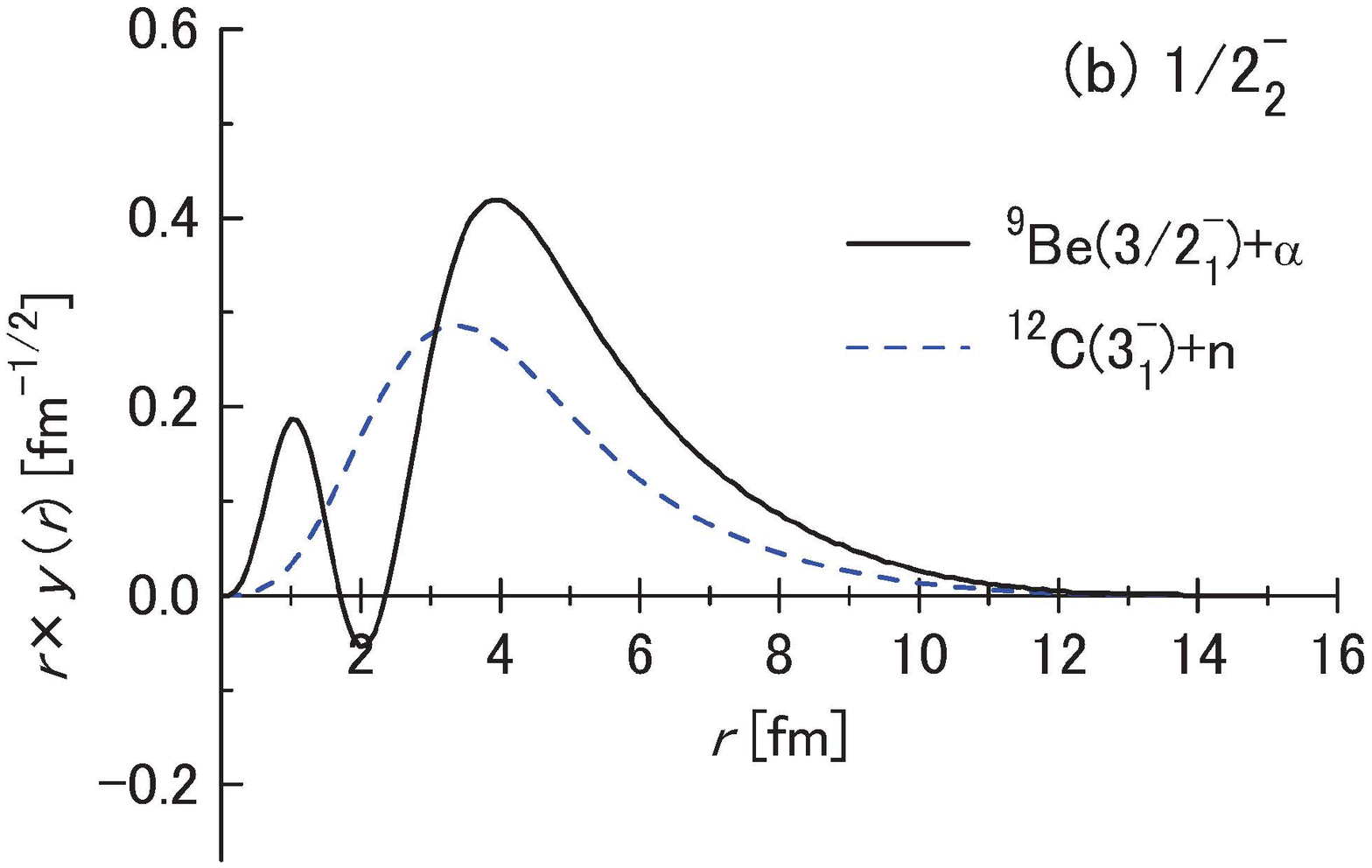}
\\ \vspace*{5mm}
\includegraphics[width=81mm]{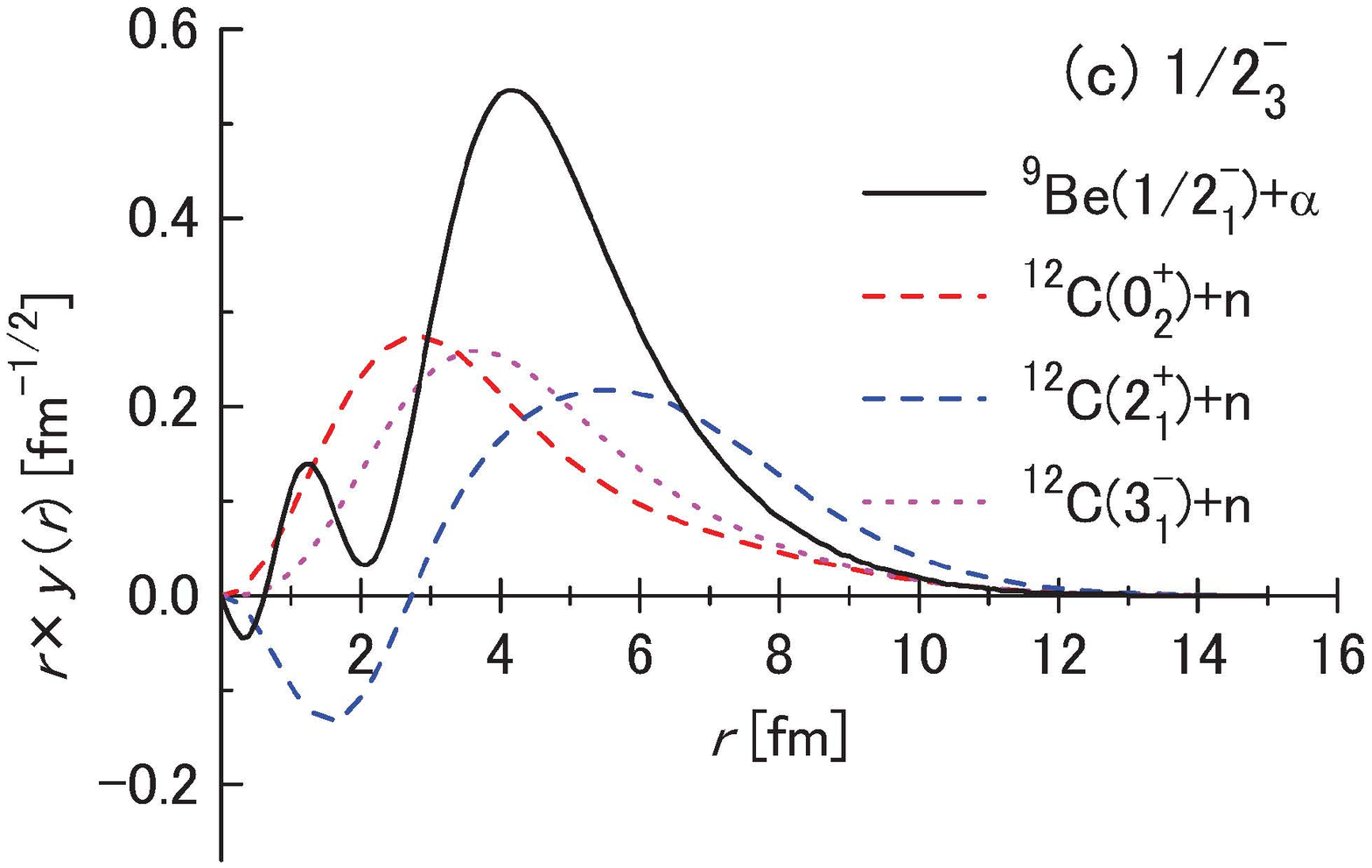}
\includegraphics[width=81mm]{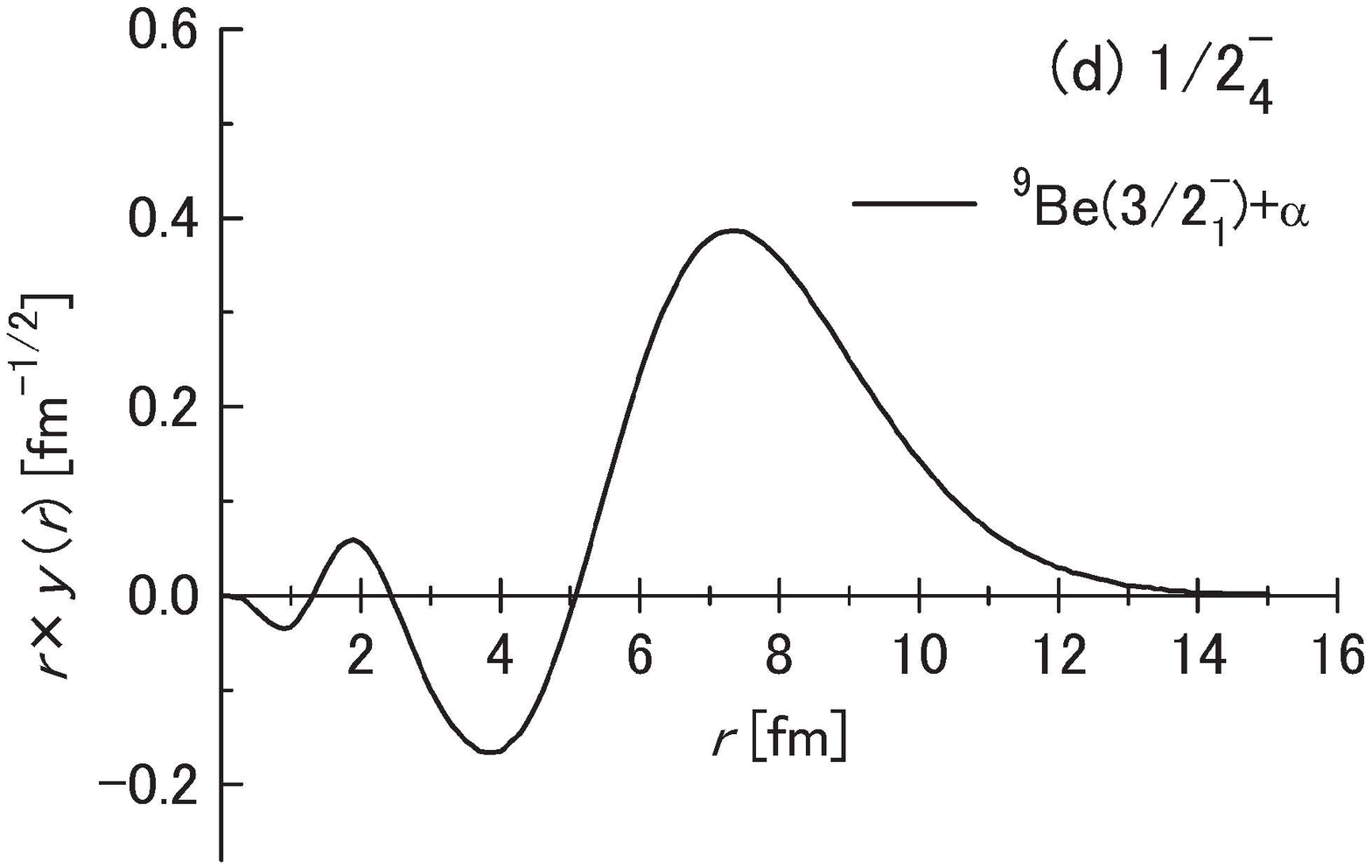}
\\ \vspace*{5mm}
\includegraphics[width=81mm]{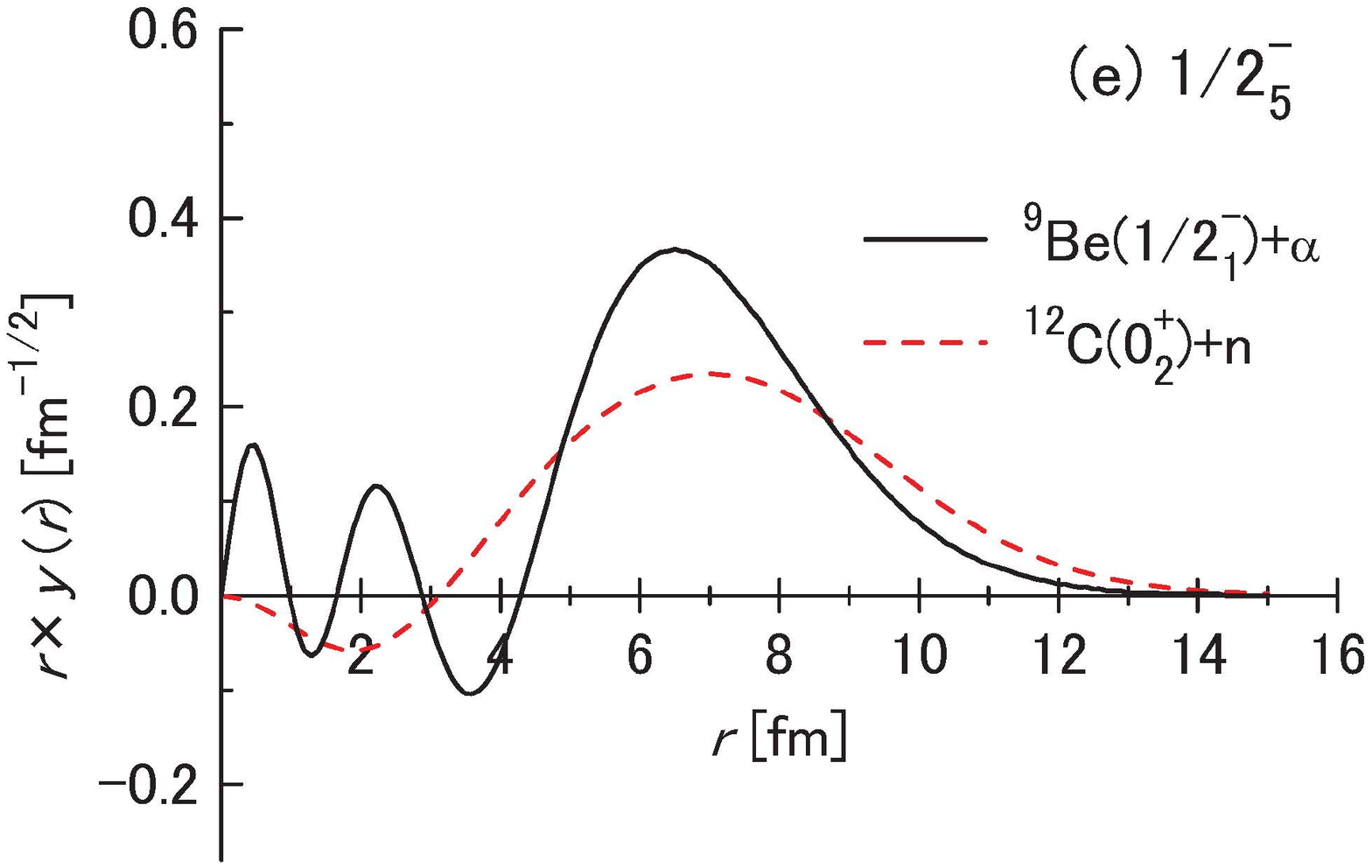}
\caption{
(Color online) Overlap amplitudes of the $^{12}$C+$n$ channels and $^{9}$Be+$\alpha$ channels for the five $1/2^{-}$ states of $^{13}$C defined in Eqs.~(\ref{eq:RWA_neutron}) and (\ref{eq:RWA_alpha}). In the panels we present only the overlap amplitudes with the $S^2$ factor larger than $0.2$ in Fig.~\ref{fig:s2_factors_12_minus_13c}.
}
\label{fig:rwa_12_minus_13c}
\end{center}
\end{figure}

The spectroscopic factors and overlap amplitudes of the $^{12}$C+$n$ and $^{9}$Be+$\alpha$ channels defined in Eqs.~(\ref{eq:RWA_neutron}) $\sim$ (\ref{eq:s2-factor}) are useful to see the structure of the \textcolor{black}{$1/2^-_1$} state.
Their results are shown in Figs.~\ref{fig:s2_factors_12_minus_13c}(a) and \ref{fig:rwa_12_minus_13c}(a). 
The values of the spectroscopic factors and radial behaviors of the overlap amplitudes of the $1/2^-_1$ state can be explained \textcolor{black}{qualitatively} by the SU(3) nature of the state.
The fact that the spectroscopic factors of the $^9$Be+$\alpha$ channels are non-zero ($S^2\sim0.2$) indicates that the ground state has not only the mean-field degree of freedom but also $\alpha$-cluster degree of freedom. 
This will be discussed in detail in Sec.~\ref{subsec_monopole}. 

The ground-state wave function of $^{13}$C can also be studied by calculating the Gamow-Teller transition rates $B({\rm GT})$ between the $^{13}$C ground state ($1/2^-_1$) and $^{13}$N states ($1/2^-_1,3/2^-_1$) in Eq.~(\ref{eq:GT}), together with the E1 transition rate $B(E1)$ between the ground state and first $1/2^+$ state of $^{13}$C in Eq.~(\ref{eq:E1}).
The calculated $B({\rm GT})$ values together with the experimental data are given as follows: $B^{\rm cal}({\rm GT})=0.332$ vs.~$B^{\rm exp}({\rm GT})=0.207\pm0.002$ for the transition from the $^{13}$C($1/2^{-}_1$) state to $^{13}$N($1/2^{-}_1$), $B^{\rm cal}({\rm GT})=1.27$ vs.~$B^{\rm exp}({\rm GT})=1.37\pm0.07$ from  $^{13}$C($1/2^{-}_1$) to $^{13}$N($3/2^-_1$). 
The calculated results are in agreement with the experimental data within a factor of $1.5$. 
On the other hand, the calculated value of $B({\rm E1}:{1/2^{-}_{1}} \rightarrow {1/2^+_1})$ is $2.0 \times 10^{-3}$~fm$^2$ in the present study, while the experimental value is $14 \times 10^{-3}$~fm$^2$.
This enhanced E1 transition rate has been pointed out by Millener et al.~\cite{millener89}, where the result of the shell model calculation is $B({\rm E1})=9.1 \times 10^{-3}$~fm$^2$, which is about two-third of the experimental value.

\begin{table}[t]
\begin{center}
\caption{Excitation energies ($E_{x}$), r.m.s.~radii ({$R$}), $C0$ transition matrix elements [$M({\rm C0})$], isoscalar monopole transition matrix elements [${\mathcal M}({\rm IS})$] of the excited $1/2^-$ states in $^{13}$C obtained by the $3\alpha+n$ OCM calculation. 
The experimental data are taken from Refs.~\cite{ajzenberg93,wittwer69} and from Ref.~\cite{kawabata08} for the $1/2^-_4$ state.  
The finite size effects of $\alpha$ particle taken into account in estimating {$R$} with the $3\alpha+n$ OCM (see Ref.~\cite{yamada05} for details). }
\label{tab:12minus_ocm}
\begin{tabular}{ccccccccc}
\hline\hline
     & \multicolumn{3}{c}{Experiment} &  &  \multicolumn{4}{c}{$3\alpha+n$ OCM} \\
      & \hspace{2mm}{$E_{x}$~[MeV]}\hspace{2mm} & \hspace{2mm}{{$R$~[fm]}}\hspace{2mm} & \hspace{2mm}$M({\rm C0})$~[fm$^{2}$]\hspace{2mm} &  \hspace*{1mm} & \hspace{2mm}{$E_{x}$~[MeV]}\hspace{2mm} & \hspace{2mm}{{$R$}}~[fm]\hspace{2mm} & \hspace{2mm}{$M({\rm C0})$}~[fm$^2$]\hspace{2mm} & \hspace{1mm}{${\mathcal M}({\rm IS})$}~[fm$^2$]\hspace{1mm} \\
\hline
\hspace{2mm}$1/2_1^-$\hspace{2mm} & $ \ 0.00 $  & $2.4628$ &  & & 0.0 &  $2.4$  &  & \\
$1/2_2^-$ & \ 8.86 & & $2.09\pm0.38$   & & 11.7 & 3.0 & 4.4 & 9.8 \\
$1/2_3^-$ & 11.08  & & $2.62\pm0.26$   & & 12.1 & 3.1 & 3.0 & 8.3 \\
$1/2_4^-$ & 12.5~  & & ${\rm No~data}$ & & 15.5 & 4.0 & 1.0 & 2.0 \\
$1/2_5^-$ & 14.39\  & & ${\rm No~data}$ & & 16.6 & 3.7 & 2.0 & 3.3 \\
\hline\hline
\end{tabular}
\end{center}
\end{table}

The four excited $1/2^{-}$ states, $1/2^{-}_{2}$, $1/2^{-}_{3}$, $1/2^{-}_{4}$, and $1/2^{-}_{5}$, have the larger nuclear radii ($3.0$, $3.1$, $4.0$ and $3.7$ fm, respectively) than that of the ground state (see Table~\ref{tab:12minus_ocm}).
As for the $1/2^{-}_2$ state, the dominant spectroscopic factor is the $^{9}$Be($3/2^{-}_{1}({\rm g.s})$)+$\alpha$ channel (see Fig.~\ref{fig:s2_factors_12_minus_13c}(b)), although one sees non-negligible contributions from the $^{12}$C($3^-_1$)+$n$ channel etc.
From the analyses of the overlap amplitudes of the $1/2^-_2$ state shown in Fig.~\ref{fig:rwa_12_minus_13c}(b), the relative wave function between $^{9}$Be and $\alpha$ in the $^{9}$Be($3/2^{-}_{1}({\rm g.s})$)+$\alpha$ channel, drawn by the real line, has a two-node $D$-wave radial behavior ($2D$),
\textcolor{black}{
where its magnitude is rather small in the inside region ($r \leq 2.5$~fm), and the magnitude of the maximum peak around $r=4$~fm is about twice larger than that of the amplitude in the inside region ($r \leq 2.5$~fm), and the tail is extended to the region of $r \sim 12$~fm.  
It is noted that its radial behavior is much different from the $2D$-type harmonic oscillator wave function (h.o.w.f.) with the nucleon size parameter $b=\sqrt{\hbar/M\omega}$ ($M$:~nucleon mass), in which the magnitude of the three peaks are almost  the same and its h.o.w.f.~has no longer tail compared with the real line in Fig.~\ref{fig:rwa_12_minus_13c}(b).
}
The reason why the relative orbital momentum is $\ell_{94}=2$ comes from the angular momentum coupling between the angular momentum of $^9$Be ($J_{9}=3/2^{-}$) and $^9$Be$-\alpha$ relative orbital angular momentum ($\ell_{94}=2$),  $\vc{J}=\vc{J_9}+\vc{\ell_{94}}$.
These results indicate that the $1/2^-_2$ state has the main configuration of $^{9}$Be($3/2^-_1$)+$\alpha$, in which the $\alpha$ cluster orbits around the $^{9}$Be({\rm g.s}) core with $2D$ orbit, \textcolor{black}{that is, a $^9$Be+$\alpha$ molecular structure is formed.}
Since the ground state ($1/2^-_1$) has a $^9$Be($3/2^-_1$)+$\alpha$ cluster degree of freedom with one-node $D$-wave ($1D$) radial behavior (see Fig.~\ref{fig:rwa_12_minus_13c}(a)), the $1/2^-_2$ state can be regarded 
\textcolor{black}{qualitatively} as an excitation of the relative motion between $^9$Be and $\alpha$ in the ground state, i.e., from $1D$ to $2D$.
\textcolor{black}{
However, one should note that the excitation from $1D$ to $2D$ is not merely the  $2\hbar\omega$ excitation in the sense of the shell model, because the ground state has significant $\alpha$-type ground-state correlation and the radial behavior of the $^9$Be($3/2^-_1$)+$\alpha$ relative wave function in the $1/2^-_2$ state qualitatively is much different from the $2D$-type h.o.w.f., as mentioned above.} 




On the other hand, the $1/2^{-}_{3}$ state has the dominant spectroscopic factor of the $^{9}$Be($1/2^-$)+$\alpha$ channel ($S^2=0.70$), and its overlap amplitude has a maximum peak in the outmost ($r \sim 4$~fm).
Thus, this state mainly has the $^{9}$Be($1/2^-$)+$\alpha$ cluster structure, where the $\alpha$ particle occupies an $S$ orbit with $3S$-like oscillatory behavior around the $^{9}$Be($1/2^-$) core (see Fig.~\ref{fig:rwa_12_minus_13c}(c)),  although the $1/2^-_3$ state has non-negligible components of the $^{12}$C($0^{+}_{2},2^{+}_{1},3^{-}_{1}$)+$n$ channels with $S^2 \sim 0.2$.
The ground state ($1/2^{-}_{1}$) has also the $^9$Be($1/2^-_1$)+$\alpha$ cluster degree of freedom with $2S$ behavior (see Fig.~\ref{fig:rwa_12_minus_13c}(a)).
Thus, the $1/2^{-}_{3}$ state can be regarded as an excitation of the relative motion between $^9$Be($1/2^-$) and $\alpha$ in the ground state, i.e., from $2S$ to $3S$.

As for the $1/2^{-}_4$ and $1/2^{-}_5$ states, the analyses of the spectroscopic factors and overlap amplitudes indicate that the $1/2^{-}_4$ state has a $^9$Be($3/2^-$)+$\alpha$ cluster structure with higher nodal behavior ($3D$), where its overlap amplitude has a maximum peak around $r=7$~fm. On the other hand, the $1/2^{-}_5$ state has a $^9$Be($1/2^-$)+$\alpha$ cluster structure with higher nodal behavior ($4S$), where its overlap amplitude has a maximum peak around $r=6$~fm, although one sees non-negligible component of the $^{12}$C($0^+_2$)+$n$ channel ($S^2 \sim 0.2$). 

Here it is interesting to clarify the reasons why the  $^9$Be+$\alpha$ cluster states together with their higher nodal states come out in the excited $1/2^-$ states, and the $1/2^-$ state with the dominant configuration of $^{12}$C(Hoyle)+$n$  does not appear in the present study.
As mentioned in Secs.~\ref{sec:introduction} and \ref{subsec:ocm}, the $3\alpha$ OCM, the model space of which is the subspace of the present $3\alpha+n$ model space, reproduces well the low-lying structure of $^{12}$C including the $0^+_3$ and $0^+_4$ states etc.~, which have been recently observed above the Hoyle state (see Fig.~\ref{fig:2}).
According to the $3\alpha$ OCM analyses, the Hoyle state is characterized by a dominant $3\alpha$-gas-like structure, where the three $\alpha$ clusters move gas-likely, only avoiding mutual overlap due to the Pauli blocking effect, although the Hoyle state has a non-negligible $^8$Be+$\alpha$ correlation.
On the other hand, the $0^+_3$ and $0^+_4$ states of $^{12}$C have the $^8$Be($0^+$)+$\alpha$ structure with higher nodal behavior and linear-chain-like structure with the dominant configuration of $^8$Be($2^+$)+$\alpha$, respectively.
With these facts in mind we will first see what kinds of structures appear in the $1/2^-$ state of $^{13}$C, when an extra neutron is added into the Hoyle state.
In this case the extra neutron should be $P$-wave with respect to the center of mass of the $3\alpha$ system, because of the parity conservation and angular momentum coupling. 
As mentioned in Sec.~\ref{subsec:ocm}, the $\alpha$-$n$ force has the strong parity dependence:~The odd-parity $\alpha$-$n$ force is significantly attractive to produce the $P$-wave resonant states ($J^\pi=3/2^-$ and $1/2^-$) in $^5$He ($\alpha$+$n$) system, while the even-parity one is weakly repulsive.
This attractive $P$-wave $\alpha$-$n$ force makes a bound state in $^9$Be ($J^\pi=3/2^-$) with respect to the $2\alpha$+$n$ threshold, having a loosely coupled $2\alpha$+$n$ structure, while the $J^\pi=1/2^+$ state appears a quasi-bound state just above the three-body threshold (see Fig.~\ref{fig:3}). 
Thus, with the addition of the $P$-wave extra neutron into the Hoyle state, the attractive $\alpha$-$n$ force reduces the size of the Hoyle state with the $3\alpha$ gas-like structure to reinforce the $^8$Be+$\alpha$ correlation in the $3\alpha+n$ system.
The extra neutron can be bound with the $^8$Be part and then the $^9$Be+$\alpha$  cluster structures are produced in excited $^{13}$C states.
In fact the $1/2^-_2$ and $1/2^-_3$ states have the $^9$Be($3/2^-$)+$\alpha$ and $^9$Be($1/2^-$)+$\alpha$ cluster structures in the present calculation, where the $^9$Be($3/2^-$) and $^9$Be($1/2^-$) states are the spin-doublet states each other.

On the other hand, the $1/2^-_4$ and $1/2^-_5$ states in the present calculation come out as the higher nodal states of $^9$Be($3/2^-$)+$\alpha$ and $^9$Be($1/2^-$)+$\alpha$, respectively, and the energy difference between the $1/2^-_4$ and $1/2^-_2$ states ($1/2^-_5$ and $1/2^-_3$) is as small as a few MeV (see Fig.~\ref{fig:4}).
These situations are similar to that in $^{12}$C. 
In fact the $0^+_3$ state of $^{12}$C has the $^8$Be($0^+$)+$\alpha$ structure with higher nodal behavior, which appears by a few MeV above the Hoyle state (see Fig.~\ref{fig:2}).
Thus, the $1/2^-_{4,5}$ states can be regarded as the counterpart of the $0^+_3$ state of $^{12}$C.

\subsubsection{Monopole excitations}\label{subsec_monopole}

The calculated {\rm C0} transition matrix elements, $M({\rm C0})$, defined in Eq.~(\ref{eq:me_e0}) are shown in Table~\ref{tab:12minus_ocm}. 
The $M({\rm C0})$ values of the $1/2^-_2$  and $1/2^-_3$ states are in agreement with the experimental data within a factor of 2 at most.
It is noted that the shell model calculation~\cite{millener89} is difficult to reproduce both the {\rm C0} matrix elements.  
\textcolor{black}{These facts support} that the $1/2^-_2$  and $1/2^-_3$ states have cluster structures.
There are no data for the $M({\rm C0})$ values of the $1/2^-_{4,5}$ states, although their calculated values are smaller than those of the $1/2^{-}_{2,3}$ states.

The isoscalar monopole matrix elements, ${\mathcal M}({\rm IS})$, defined in Eq.~(\ref{eq:me_is_monopole}) are also shown in Table~\ref{tab:12minus_ocm}. 
The calculated results of ${\mathcal M}({\rm IS})$ are about twice larger than those of $M({\rm C0})$ for the four excited $1/2^{-}_{2,3,4,5}$ states. 
The reason is given as follows:~In the present four-body $3\alpha$+$n$ model, the $M({\rm C0})$ in Eq.~(\ref{eq:me_e0}) has a relation with the isoscalar monopole matrix element ${\mathcal M}({\rm IS})$ in Eq.~(\ref{eq:me_is_monopole}) as follows:~
\begin{eqnarray}
&&{M({\rm C0},{1/2^{-}_{n}}-{1/2^{-}_{1}})} \nonumber \\
&& \hspace*{10mm} = \frac{1}{2} \times {{\mathcal M}({\rm IS},{1/2^{-}_{n}}-{1/2^{-}_{1}})}
    - \frac{1}{2} \times {\langle  {\tilde{\Phi}_{1/2^{-}_{n}}({^{13}{\rm C}})} |   (\vc{r}_{n} - \vc{R}_{\rm cm} )^2 | {\tilde{\Phi}_{1/2^{-}_{1}}({^{13}{\rm C}})} \rangle },
\label{eq:me_e0_is}
\end{eqnarray}
where $\vc{r}_{n}$ represents the coordinate of the extra neutron in the $3\alpha$+$n$ model.
The contribution from the second term in the right hand side in Eq.~(\ref{eq:me_e0_is}) is much smaller than that from the first term for the four excited $1/2^-_{2,3,4,5}$ states in the present calculation.
Thus, the relation of $M({\rm C0}) \sim (1/2) \times {\mathcal M}({\rm IS})$ is realized.
The total value of the energy weighted sum rule (EWSR) for the isoscalar monopole transition in $^{13}$C is given as ${\frac{2\hbar^2}{m} \times 13 \times R^2} = 6.5 \times 10^3$~fm$^{4}$~MeV, where $R$ denotes the nuclear radius of the ground state of $^{13}$C, and $m$ stands for the mass of nucleon.
Then, the percentages of the energy weight strength to the isoscalar monopole EWSR value are, respectively, $13\%$, $11\%$, $1\%$, and $2\%$, for the $1/2^{-}_{2,3,4,5}$ states.
The total sum of the percentages amounts to be about $25\%$, the value of which is comparable to the case of $^{16}$O ($\sim 20\%$)~\cite{yamada08_monopole,yamada12}.

The preliminary experimental data of the isoscalar monopole matrix elements in $^{13}$C have been provided by the inelastic $\alpha$ scattering experiments, performed at RCNP~\cite{kawabata08,kawabata14}; ${\mathcal M}^{\rm exp}({\rm IS})=6.1$~fm$^2$ ($6\%$), $4.2$~fm$^2$ ($3\%$), and $4.9$~fm$^2$ ($5\%$), for the $1/2^-_2$, $1/2^-_3$, and $1/2^-_4$ states, respectively, in which the values in parentheses denote the percentages of the energy weight strength to the EWSR value.
It is noted that these experimental ${\mathcal M}({\rm IS})$ values are comparable to the experimental data of the isoscalar monopole matrix element to the $0^+_2$ state (Hoyle state) in $^{12}$C, ${\mathcal M}(\rm IS)=10.7$~fm$^2$, together with that to a $^8$He+$\alpha$ cluster state in $^{12}$Be, ${\mathcal M}({\rm IS}) = 7.1 \pm 1.0$~fm$^2$, observed in a recent breakup-reaction experiment using a $^{12}$Be beam at 29 MeV/nucleon~\cite{yang14}.
The sum of the experimental percentages for the $1/2^{-}_{2,3,4}$ states of $^{13}$C amounts to about $14\%$ of EWSR.  
The calculated results of the ${\mathcal M}({\rm IS})$ values together with the sum of the percentages to the isoscalar monopole EWSR value obtained by the present $3\alpha+n$ OCM are in correspondence with the preliminary experimental ones within a factor of 2 at most.

As discussed in Sec.~\ref{subsec:structures_12minus}, the four excited $1/2^-$ states, $1/2^-_2$, $1/2^-_3$, $1/2^-_4$ and $1/2^-_5$, have the following characteristic structures, $^9$Be($3/2^-$)+$\alpha$, $^9$Be($1/2^-$)+$\alpha$, $^9$Be($3/2^-$)+$\alpha$ with higher nodal behavior, and $^9$Be($1/2^-$)+$\alpha$ with higher nodal behavior, respectively, although the ground state ($1/2^-_1$) has a shell-model-like structure. 
Here it is interesting to investigate the mechanism of why these characteristic $1/2^-_{2,3,4,5}$ states are exited by the monopole transitions from the shell-model-like ground state.

For this purpose, it is instructive to demonstrate that the SU(3) wave function, $|(0s)^4(0p)^{9}[4441](0,3)_{L=1^-}\otimes \chi_{\frac{1}{2}}(n)\rangle_{J=\frac{1}{2}^-}$, which is dominant in the ground state of $^{13}$C in the present study (see Sec.~\ref{subsub:structures_12minus}), is mathematically equivalent to a single-cluster-model wave function of $^9$Be+$\alpha$ with the total harmonic oscillator quanta $Q=9$,
\begin{eqnarray}
&&|(0s)^{4}(0p)^{9}[4441](0,3)_{L=1^-}\otimes \chi_{\frac{1}{2}}(n)\rangle_{J=\frac{1}{2}^-} \label{eq:su3_13C} \\
&& \hspace*{10mm} = N_{\frac{3}{2}} \sqrt{\frac{9!4!}{13!}} \mathcal{A}\left\{ u_{42}(\vc{r}_{94})\phi_{j={\frac{3}{2}}^{-}}(^{9}{ \rm Be}) \phi(\alpha) \right\}_{J=\frac{1}{2}^-},\label{eq:su3_13C_9Be_alpha_12minus}\\
&& \hspace*{10mm} = N_{\frac{1}{2}} \sqrt{\frac{9!4!}{13!}} \mathcal{A}\left\{ u_{40}(\vc{r}_{94})\phi_{j={\frac{1}{2}}^{-}}(^{9}{ \rm Be}) \phi(\alpha) \right\}_{J=\frac{1}{2}^-},\label{eq:su3_13C_9Be_alpha_32minus}
\label{eq:su3_13C_9Be_alpha_52minus}
\end{eqnarray}
where $N_{\frac{3}{2},\frac{1}{2}}$ are the normalization constants, and $\phi(\alpha)$ represents the internal wave function of the $\alpha$ cluster with the $(0s)^4$ configuration.
$\phi_j({^{9}}{\rm Be})$ stands for the internal wave function of $^9$Be with the angular momentum $j$, $\phi_j({^{9}}{\rm Be})=|(3,1)_{\ell}\otimes\chi_{\frac{1}{2}}(n)\rangle_j$, the spatial part of which belongs to the SU(3) representation $(\lambda,\mu)=(3,1)$ of  the $(0s)^4(0p)^5$ configuration.
The relative wave function between the $^9$Be and $\alpha$ clusters in Eqs.~(\ref{eq:su3_13C_9Be_alpha_12minus}) and (\ref{eq:su3_13C_9Be_alpha_32minus}) is described by the harmonic oscillator wave function $u_{QLM}(\vc{r}_{94})=u_{QL}(r_{94})Y_{LM}(\hat{\vc{r}}_{94})$ with the node number $n=(Q-L)/2$ and orbital angular momentum $L$, where $\vc{r}_{94}$ denotes the relative coordinate between the $^9$Be and $\alpha$ clusters.
One can prove Eqs.~(\ref{eq:su3_13C_9Be_alpha_12minus}) and (\ref{eq:su3_13C_9Be_alpha_32minus}) with help of the Bayman-Bohr theorem~\cite{bayman58}.
Equations  (\ref{eq:su3_13C_9Be_alpha_12minus}) and (\ref{eq:su3_13C_9Be_alpha_32minus}) mean that the ground state of $^{13}$C has  $^9$Be+$\alpha$ cluster degrees of freedom as well as the mean-filed degree of freedom.
We call this the dual nature of the ground state~\cite{yamada08_monopole}.
This dual nature is also realized in the ground state of $^{16}$O, $^{12}$B, and $^{11}$B etc.  

The operator of the isoscalar monopole transition of $^{13}$C, ${\mathcal O}({\rm IS},{^{13}{\rm C}}) =\sum_{i}^{13}(\vc{r}_i -\vc{R}_{\rm cm} )^2$, in Eq.~(\ref{eq:me_is_monopole}) is decomposed into the internal parts and the relative part, 
\begin{eqnarray}
{\mathcal O}({\rm IS},{^{13}{\rm C}}) = {\mathcal O}({\rm IS},{^9{\rm Be}}) + {\mathcal O}({\rm IS},{\alpha})  + \frac{9 \times 4}{13} \vc{r}_{94}^2
\label{eq:monopole_decompose}
\end{eqnarray}
where ${\mathcal O}({\rm IS},{^9{\rm Be}})$ and  ${\mathcal O}({\rm IS},{\alpha})$ stand for the isoscalar monopole operator of $^{9}$Be and $\alpha$ clusters, respectively.
The relative part, $(9\times 4/13)\vc{r}_{94}^2$, can excite the relative motion between the $^{9}$Be and $\alpha$ clusters ($2\hbar\omega$ excitation).
In the case of the isoscalar monopole excitation from the ground state described in Eq.~(\ref{eq:su3_13C}) to $^9$Be+$\alpha$ cluster states, one can easily prove that the monopole matrix elements in Eq.~(\ref{eq:me_is_monopole}) is originated from only the contribution from the relative part, $(9\times 4/13)\vc{r}_{94}^2$, in Eq.~(\ref{eq:monopole_decompose}).   
This proof is similar to the case of $^{16}$O, in which the $0^+_{2,3}$ states with the $^{12}$C($0^{+}_{1}$,$2^{+}_{1}$)+$\alpha$ cluster structures are excited by the monopole transitions from the ground state with doubly closed shell structure~\cite{yamada08_monopole,yamada12}.
Consequently the $^{9}$Be($3/2^-$,$1/2^-$)+$\alpha$ cluster degrees of  freedom embedded in the ground state are activated by the monopole operator and then the $1/2^{-}_{2,3}$ states with the $^{9}$Be($3/2^-$,$1/2^-$)+$\alpha$ cluster structures are excited by the monopole transitions. 
In fact one can see that the overlap amplitude of the $^9$Be($3/2^-$)+$\alpha$ channel in the $1/2^-_2$ state shows $2D$-like oscillatory behavior, which has one node higher than that in the ground state having $1D$-like relative motion (see Figs.~\ref{fig:rwa_12_minus_13c}(a) and (b)).  
In addition, in the $1/2^-_3$ state, the overlap amplitude of the $^9$Be($1/2^-$)+$\alpha$ channel has $3S$-like oscillatory behavior, while that in the ground state has $2S$-like one (see Figs.~\ref{fig:rwa_12_minus_13c}(a) and (c)).
\textcolor{black}{
Therefore, the $1/2^-_2$ and $1/2^-_3$ states, which have the dominant $^9$Be+$\alpha$ cluster structures, can be excited by the isoscalar monopole operator from the SU(3) $(0,3)$ state.
However, according to the present calculation, the isoscalar monopole matrix elements of the $1/2^-_{2,3}$ states from the SU(3) $(0,3)$ state in Eq.~(\ref{eq:su3_13C}), $\langle 1/2^-_{2,3} | \mathcal{O}({\rm IS},{^{13}{\rm C}}) | [(0,3)_{L=1^{-}}\otimes \chi_{\frac{1}{2}}(n)]_{J=1/2^-}  \rangle$, are less than one-third or further smaller compared with ones shown in Table~\ref{tab:12minus_ocm}, $\langle 1/2^-_{2,3} | \mathcal{O}({\rm IS},{^{13}{\rm C}}) | {1/2^-_1} \rangle$, in which the wave function of the $1/2^-_{1}$ state is one obtained by the present $3\alpha+n$ cluster model.
As mentioned in Sec.~\ref{subsub:structures_12minus}, the $1/2^-_{1}$ state has significant $\alpha$-type ground-state correlation:~The component of 
the $0\hbar\omega$ basis is $61~\%$ in the present study, and the remaining comes from the higher $\hbar\omega$ bases.
We found that the calculated isoscalar monopole matrix elements (and C0 ones) in Table~\ref{tab:12minus_ocm} dominantly comes from the coherent contribution between the $\alpha$-type ground-state correlation in the $1/2^-_1$ state and the $1/2^-_{2,3}$ wave functions having spatially developed $^9$Be+$\alpha$ cluster structures. 
}
\textcolor{black}{This} mechanism of the excitation of cluster states from the shell-model-like ground state by the monopole transition is common to the cases of $^{16}$O, $^{12}$C, $^{11}$B, and $^{12}$Be etc.~which have been discussed in Refs.~\cite{yamada08_monopole,yamada12,yamada10,ito12,ito14}.  

As for the isoscalar monopole transitions to the $1/2^-_4$ and $1/2^-_5$ states, their matrix elements are about half or one third smaller than those to the $1/2^-_2$ and $1/2^-_3$ states, although the radii of the former states are lager than those of the latter  (see Table~\ref{tab:12minus_ocm}).
The reasons are given as follows:~The structures of the $1/2^-_4$ and $1/2^-_5$ states are characterized by the higher nodal behavior in the $^9$Be($3/2^-$)-$\alpha$ and $^9$Be($1/2^-$)-$\alpha$ relative motions, respectively, i.e.~one node higher compared with the $^9$Be+$\alpha$ relative motions in the $1/2^-_2$ and $1/2^-_3$ states. 
Reflecting this higher nodal character, the maximum peak in the overlap amplitude of the $^9$Be+$\alpha$ channels in the $1/2^-_{4,5}$ states appears at as large as $r=6\sim 7$~fm, while that in the $1/2^-_{2,3}$ states (the ground state) is located at $r\sim 4$~fm ($r\sim 2.5$~fm), as shown in Fig.~\ref{fig:rwa_12_minus_13c}.  
The relative part of the isoscalar monopole operator in Eq.~(\ref{eq:monopole_decompose}) can excite the relative motion of the $^9$Be+$\alpha$ cluster degree of freedom in the ground state  by $2\hbar\omega$ with no change of the relative orbital angular momentum.
In other word this operator can mainly populate the $^9$Be+$\alpha$ cluster states having {\it one} node higher in the relative motion of $^9$Be+$\alpha$ than the ground state.   
As mentioned above, the $1/2^-_{4,5}$ states correspond to \textcolor{black}{the} $^9$Be+$\alpha$ cluster states having {\it two} nodes higher than the ground state.
Thus the $1/2^-_{4,5}$ states are not excited strongly by the monopole transition compared with the cases of the $1/2^-_{2,3}$ states.
This result indicates that the monopole transition to cluster states with larger radius such as the relevant $1/2^-_{4,5}$ states is not always stronger. 

The main feature of the $1/2^{-}_{4}$ and $1/2^{-}_{5}$ states having one node higher in the $^9$Be-$\alpha$ relative motion compared with the case of the $1/2^{-}_{2}$ and $1/2^{-}_{3}$ states, respectively, can also be verified by the strong isoscalar monopole transitions and strong {\rm C0} transitions between them. 
The calculated matrix elements are given as follows:~${\mathcal M}({\rm IS};{1/2^-_2}-{1/2^-_4})=28$~fm$^2$,  ${\mathcal M}({\rm C0};{1/2^-_2}-{1/2^-_4})=13$~fm$^2$;~${\mathcal M}({\rm IS};{1/2^-_3}-{1/2^-_5})=22$~fm$^2$,  ${\mathcal M}({\rm C0};{1/2^-_3}-{1/2^-_5})=8$~fm$^2$.
These values are about three times larger than those from the ground state ($1/2^-_1$) to the $1/2^-_{2,3}$ states (see Table~\ref{tab:12minus_ocm}).
These results support that the structures of the $1/2^-_4$ and $1/2^-_5$ states are characterized by the higher nodal behavior in the $^9$Be($3/2^-$)-$\alpha$ and $^9$Be($1/2^-$)-$\alpha$ relative motions, respectively.

\textcolor{black}{Here} we make a comment on the interpretations of the $1/2^-_2$ 8.86-MeV state by the present $\alpha$ cluster model and shell model~\cite{millener89,cohen65}.
According to the shell-model calculation~\cite{millener89}, the $1/2^{-}_{2}$ state is interpreted as the $p$-shell one with the dominant configuration of SU(3)$[f](\lambda,\mu)=[432](1,1)$, based on the experiments with one-nucleon and two-nucleon pickup reactions~\cite{fleming68,hinterberger68} and the $0\hbar\omega$ shell-model calculation~\cite{cohen65}, although the experimental C0 matrix element of the $1/2^-_2$ state as well as the $1/2^-_3$ state is not reproduced at all in their calculations, and there are no papers reproducing them with the $(0+2)\hbar\omega$ shell model calculations as far as we know. 
On the other hand, the present $\alpha$ cluster model for the first time has reasonably reproduced the experimental C0 matrix elements (and the isoscalar monopole matrix elements obtained by the $^{13}$C$(\alpha,\alpha')$ reaction).
However, the C0 matrix element of the $1/2^-_2$ state is overestimated by a factor of about 2 compared with the experimental data, while that of the $1/2^-_3$ state is only $1.1$ time larger (see Table~\ref{tab:12minus_ocm}).

\textcolor{black}{
In addition to the C0 matrix elements as well as the IS monopole matrix elements, another interesting experimental information on the $1/2^{-}_{2,3}$ states is the M1 transition strengths to the ground state.
According to the shell model calculation~\cite{millener89,cohen65}, the M1 transition strength from the $1/2^{-}_{2}$ state ($8.86$-MeV) is about twice larger than the experimental data ($0.23$ W.u.), while those from the $11.75$-MeV ($3/2^-_2$,~$T=1/2$) and $15.11$-MeV ($3/2^-$,~$T=3/2$) are reproduced well within a factor of about $1.1$.
It is reminded that the shell model calculation claims that the $8.86$-MeV, $11.75$-MeV, and $15.11$-MeV levels are mainly $p$-shell states with the $[432]$ symmetry.
Only the M1 strength from the $1/2^{-}_{2}$ state ($8.86$-MeV) is overestimated by a factor of about $2$.
On the other hand, the experimental M1 strength from the $1/2^{-}_{3}$ state ($11.08$-MeV) is much weaker ($0.036$ W.u.), and there are no theoretical calculations for it, as far as we know.  
}

From these \textcolor{black}{experimental data and theoretical analyses of the $1/2^-$ states}, the present overestimation of the C0 matrix element of the $1/2^-_2$ state may indicate that the $1/2^-_2$ state is the admixture of the $^9$Be($3/2^-$)+$\alpha$ cluster component and the $0\hbar\omega$ shell-model component with the $[432]$ symmetry.
It is considered that the C0 matrix element of this admixture will be reduced compared with one obtained by the present $\alpha$ cluster \textcolor{black}{model}, and the $\alpha$ cluster component in this admixture is dominantly responsible for the monopole matrix element and the $0\hbar\omega$ shell-model component with the $[432]$ symmetry is mainly attributed to the one- and two-nucleon pickup reactions~\cite{millener89,cohen65}.
\textcolor{black}{
This admixture may also reduce the M1 transition strength of the $1/2^-_2$ state, which is overestimated in the shell model calculation~\cite{millener89,cohen65}. 
}
Although further theoretical analyses are needed to solve \textcolor{black}{these problems}, it is believed that the present study has indicated the importance of the $\alpha$-clustering aspects in the $1/2^-_2$ state.

\subsection{$1/2^{+}$ states}\label{sub:structures_12plus}

\begin{table}[t]
\begin{center}
\caption{Excitation energies ($E_{x}$) and r.m.s.~radii ({$R$}) of the $1/2^+$ states in $^{13}$C obtained by the $3\alpha+n$ OCM calculation. 
The experimental data are taken from Ref.~\cite{ajzenberg93}.  
The calculated C0 matrix elements $M({\rm C0})$ and isoscalar monopole matrix elements ${\mathcal M}({\rm IS})$ from the $1/2^+_1$ state to the $n$th excited $1/2^+$ state are shown for reference.
}
\label{tab:12plus_ocm}
\begin{tabular}{ccccccccc}
\hline\hline
     & \multicolumn{1}{c}{Experiment} &  &  \multicolumn{4}{c}{$3\alpha+n$ OCM} \\
      & \hspace{2mm}{$E_{x}$~[MeV]}\hspace{2mm}  &  \hspace*{10mm} & \hspace{2mm}{$E_{x}$~[MeV]}\hspace{2mm} & \hspace{2mm}{{$R$}}~[fm]\hspace{2mm}   & \hspace{2mm}{$M({\rm C0})$}~[fm$^2$]\hspace{2mm} & \hspace{1mm}{${\mathcal M}({\rm IS})$}~[fm$^2$]\hspace{1mm} \\
\hline
\hspace{2mm}$1/2_1^+$\hspace{2mm} & \ 3.089   & & \ 3.0  & 2.6  & & &\\
$1/2_2^+$ & 10.996                                          & & 11.7 & 3.2 & 4.9 & 11.0 \\
$1/2_3^+$ & 12.14\                                           & & 12.5 & 3.1 & 4.1 & 8.0 \\
$1/2_4^+$ &                                                   & & 14.2 & 4.0 & 2.9 & \ 4.6 \\
$1/2_5^+$ &                                                   & & 14.9  & 4.3 & 0.6 & \ 2.1 \\
\hline\hline
\end{tabular}
\end{center}
\end{table}

\begin{figure}[t]
\begin{center}
\includegraphics[width=53mm]{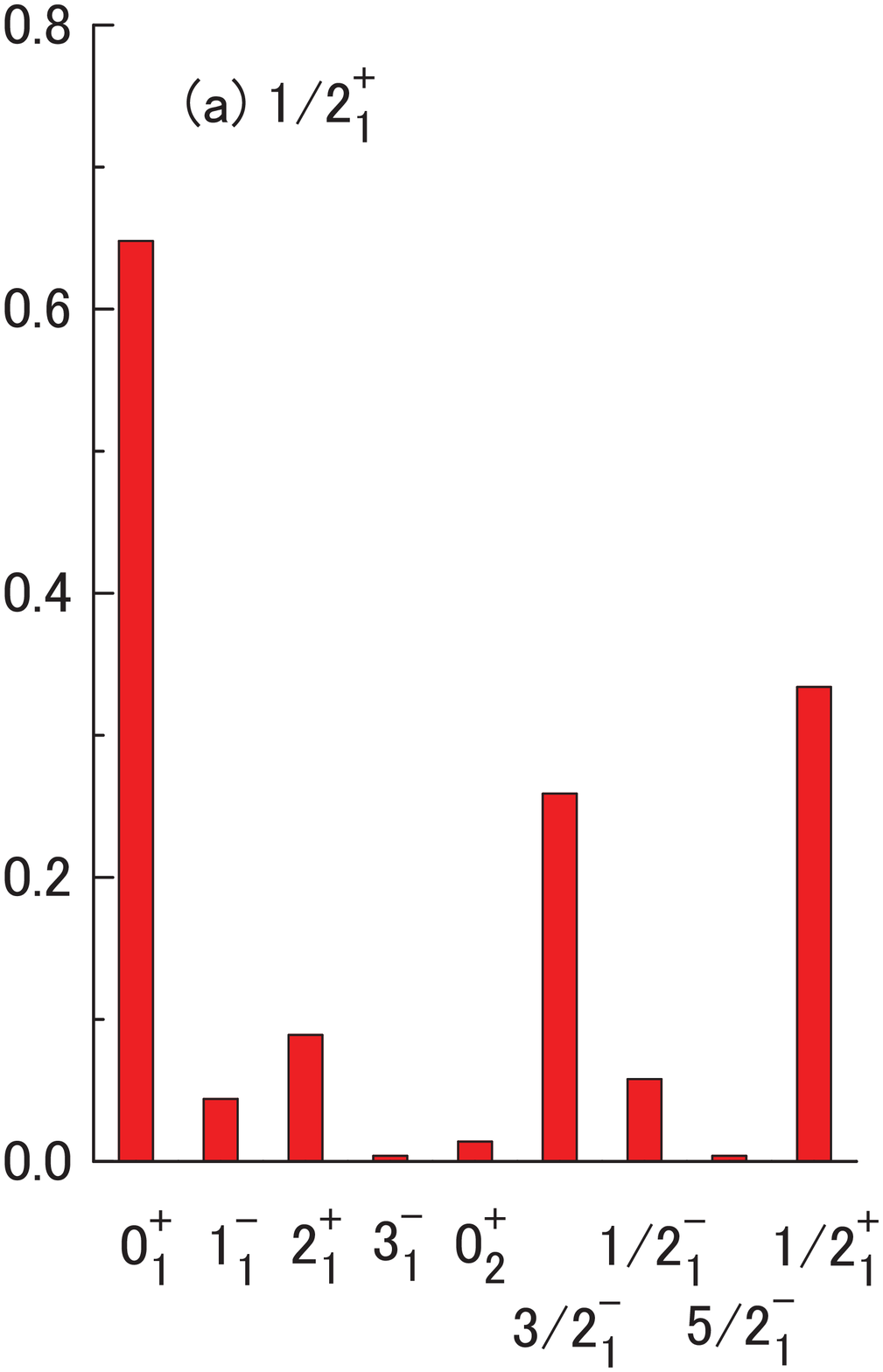}
\includegraphics[width=53mm]{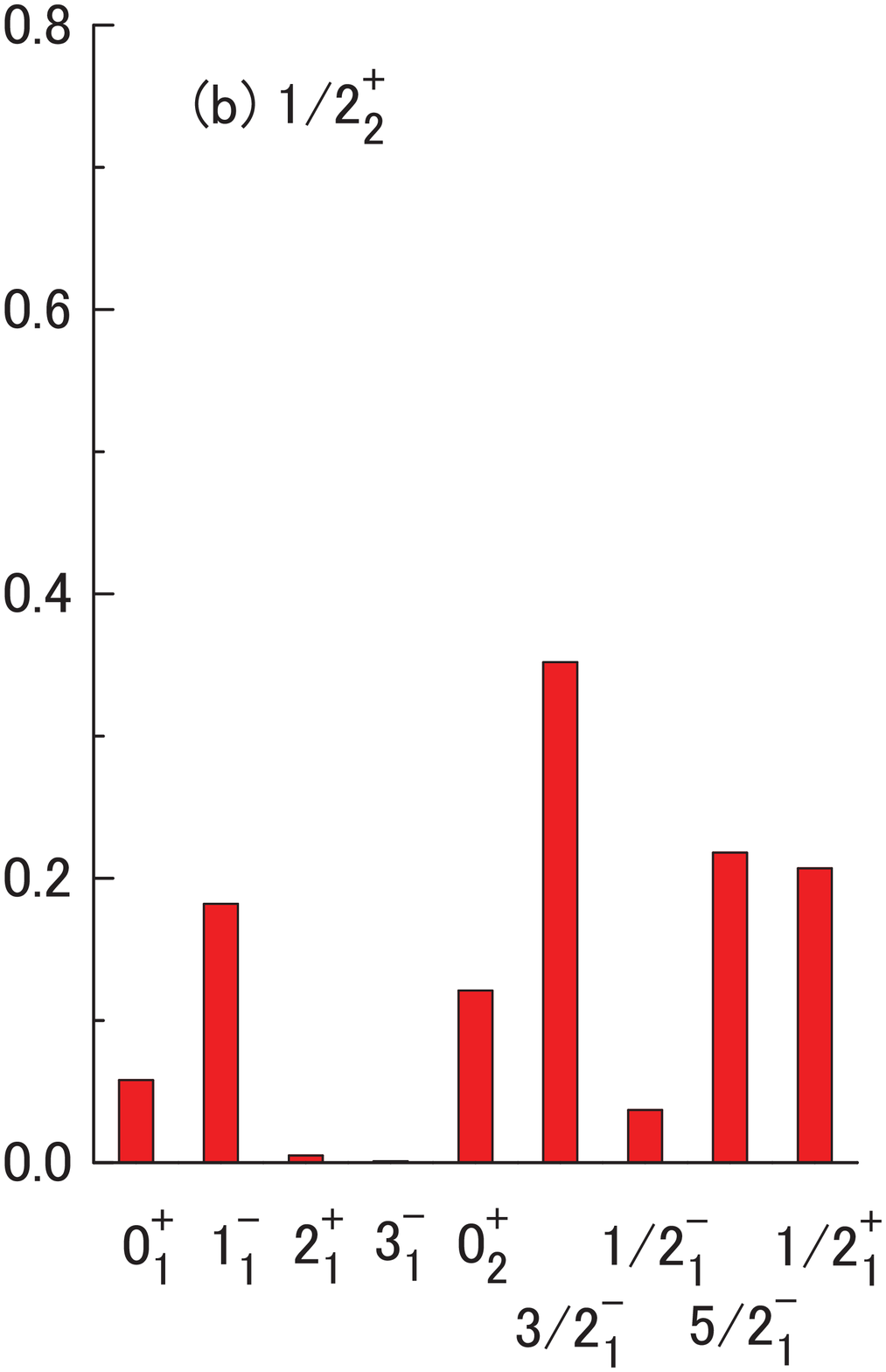}
\includegraphics[width=53mm]{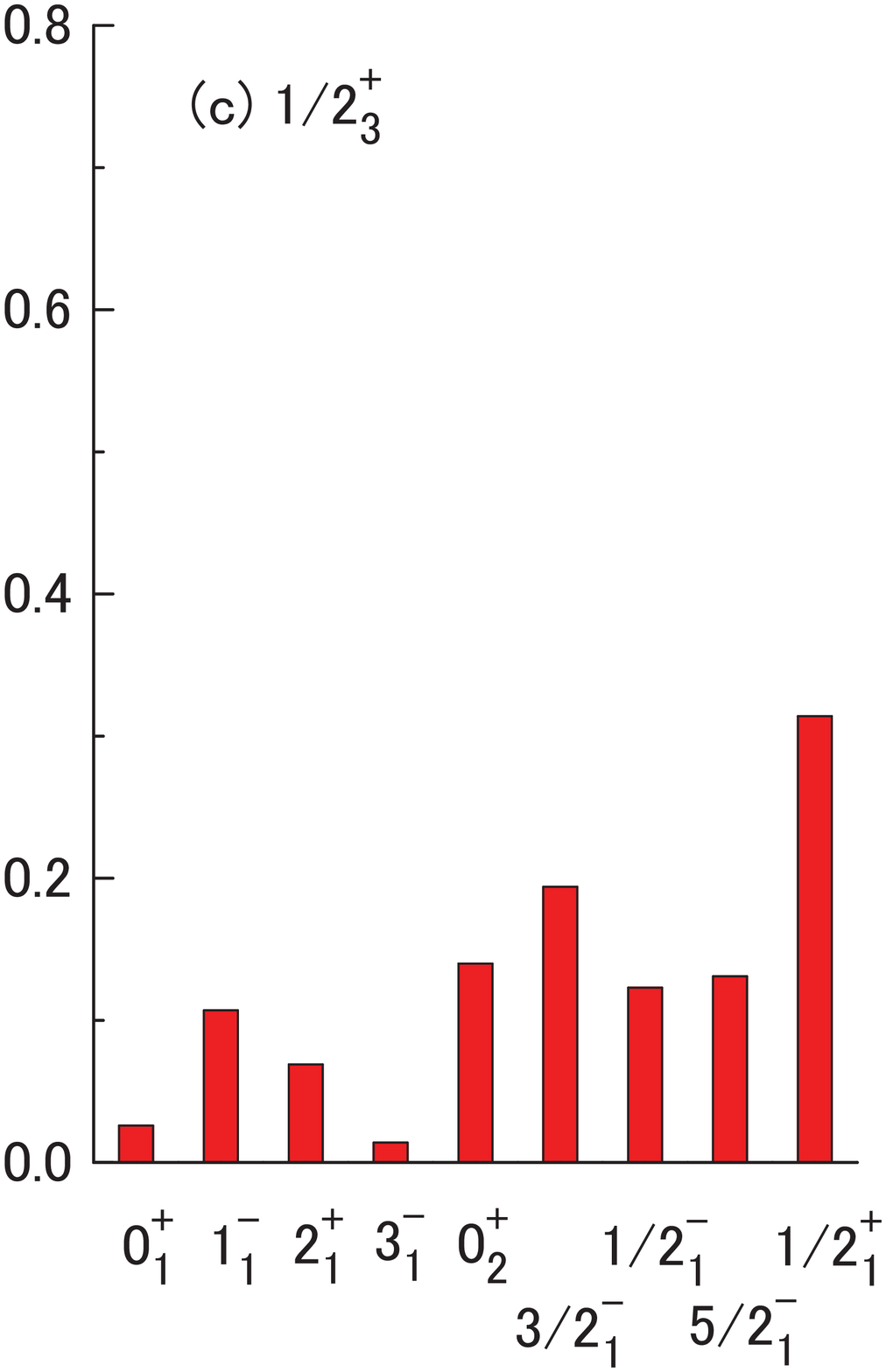}
\\ \vspace*{5mm}
\includegraphics[width=53mm]{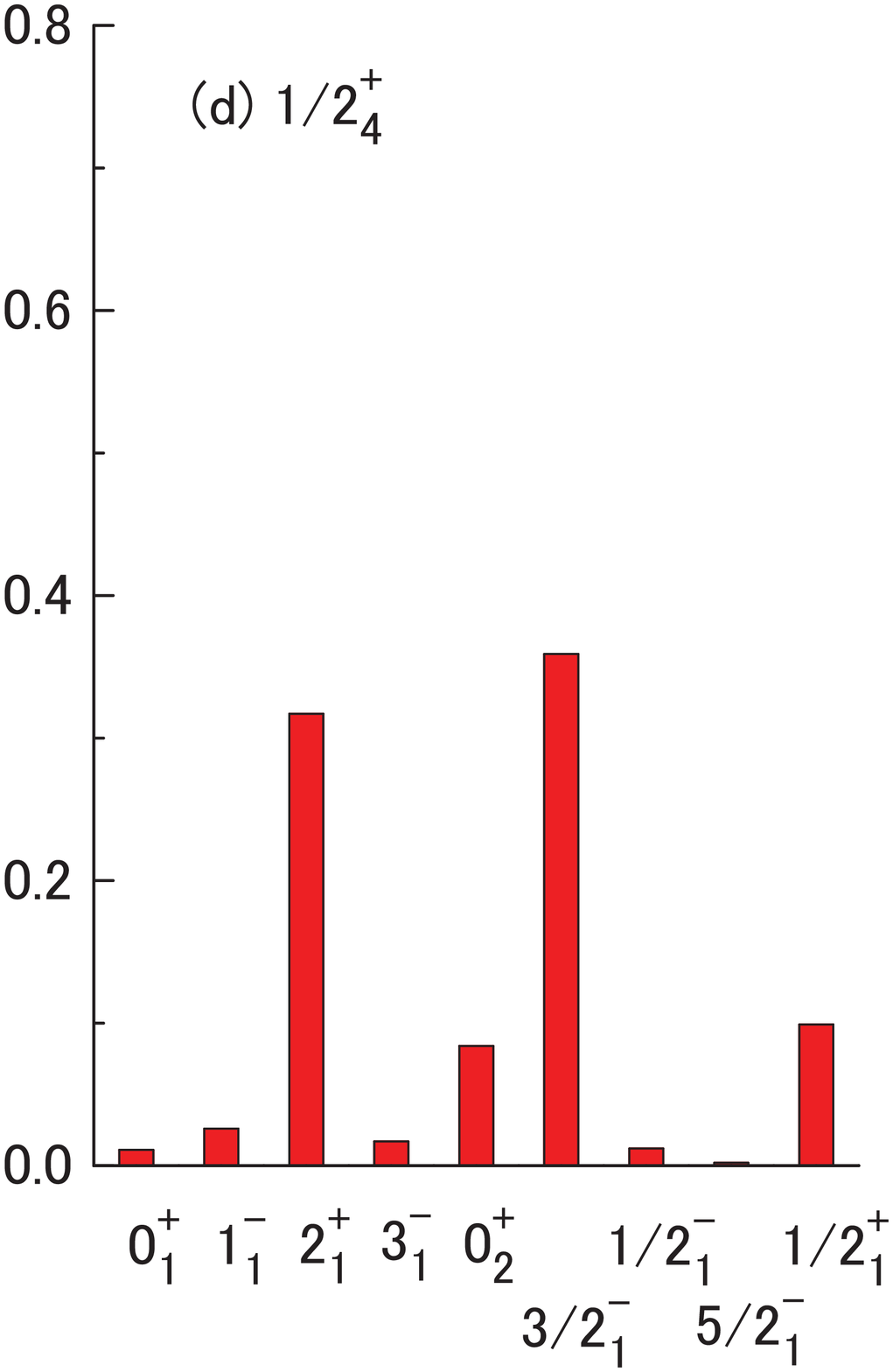}
\includegraphics[width=53mm]{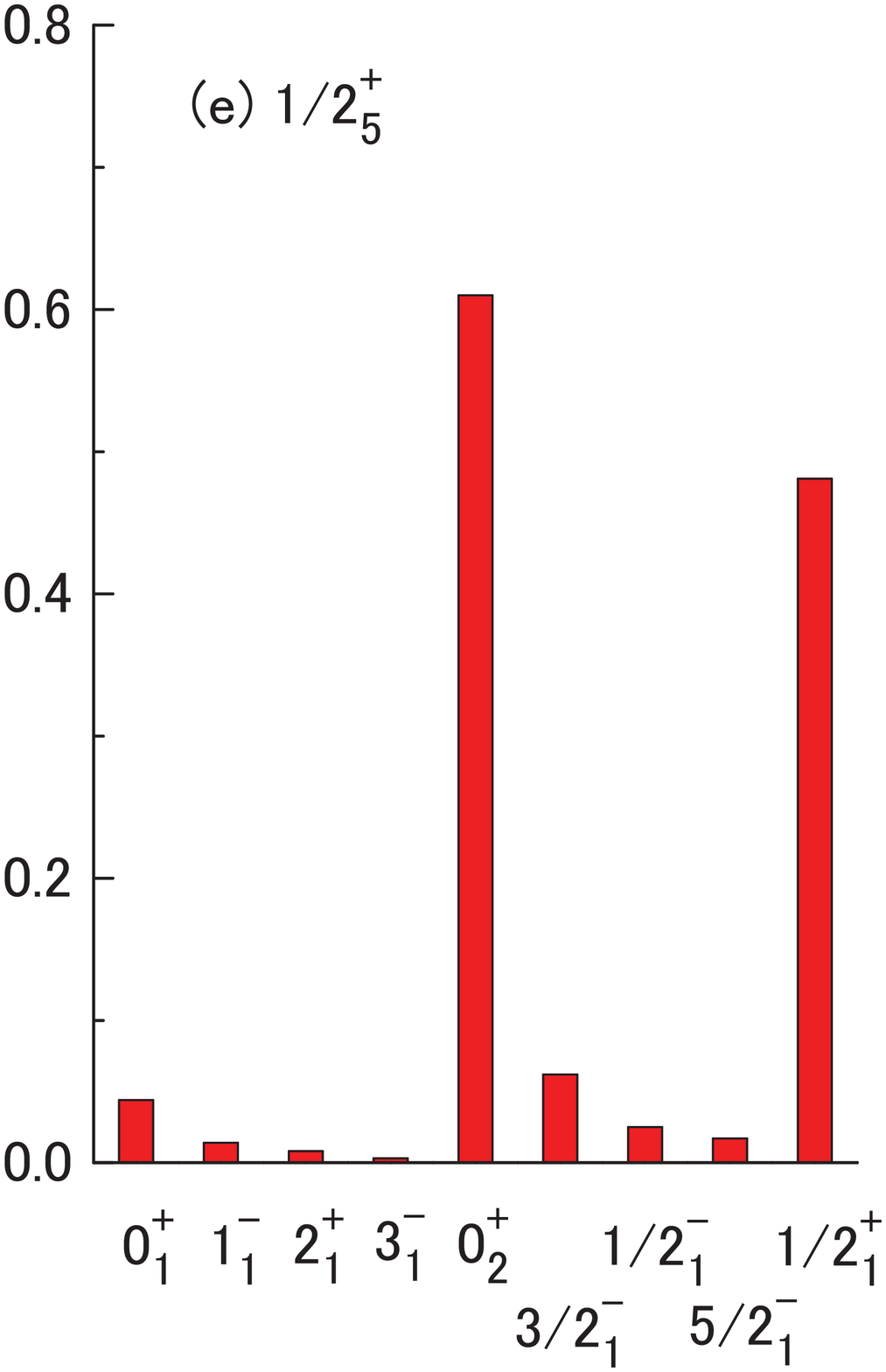}
\caption{(Color online) {Spectroscopic factors $S^{2}_{L}$} of the $^{12}$C$(J^{\pi}_C)$+$n$ channels ($J^{\pi}_C=0^{+}_{1},1^{-}_{1},2^{+}_{1},3^{-}_{1},0^{+}_{2}$) and $^{9}$Be($J^{\pi}_9$)+$\alpha$ ($J^{\pi}_9=3/2^{-}_1,1/2^{-}_1,5/2^{-}_1,1/2^{+}_1$) in the five $1/2^{+}$ states of $^{13}$C {defined in Eq.~(\ref{eq:s2-factor})}.}
\label{fig:s2_factors_12_plus_13c}
\end{center}
\end{figure}

\begin{figure}[t]
\begin{center}
\includegraphics[width=81mm]{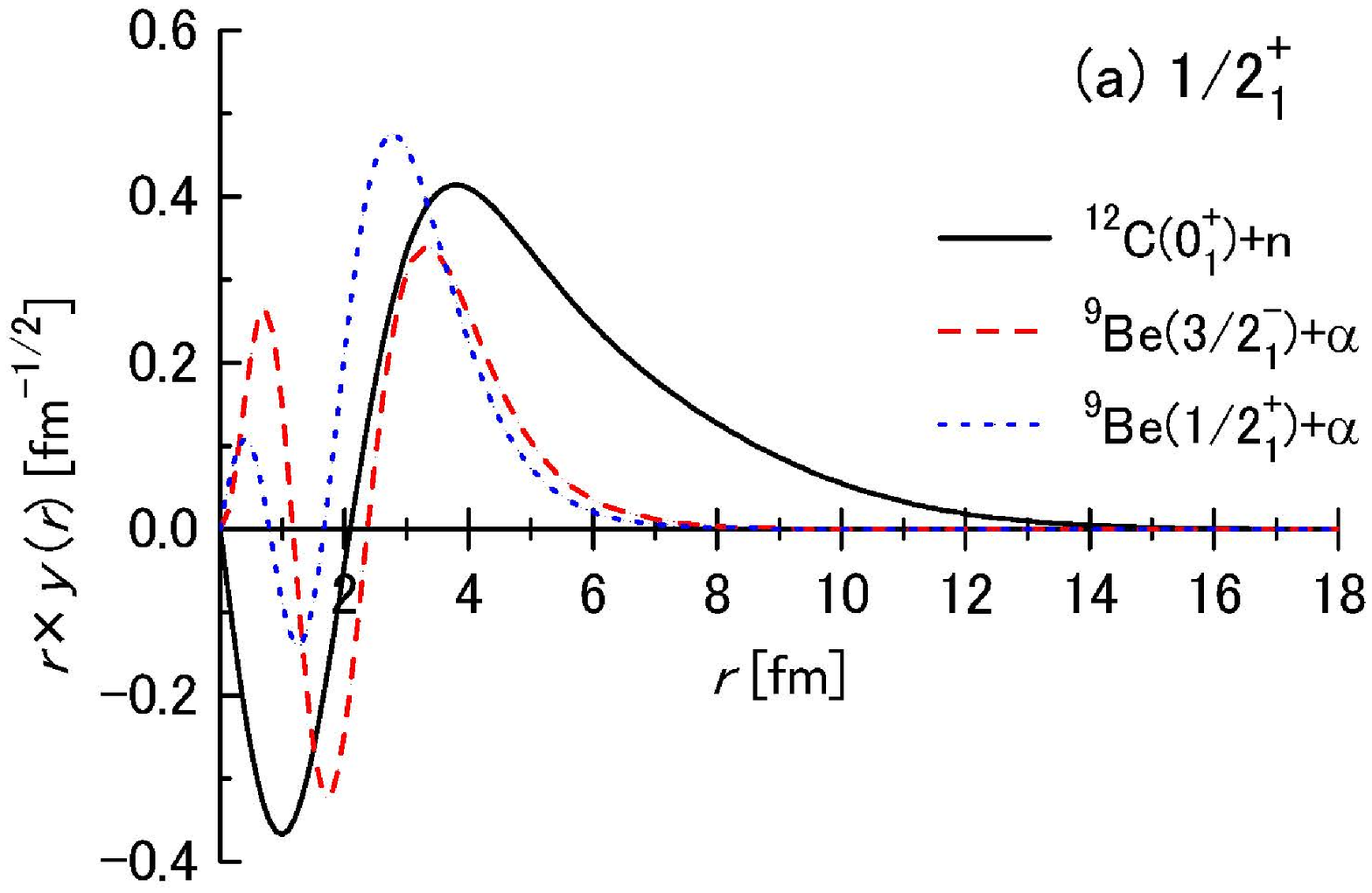}
\includegraphics[width=81mm]{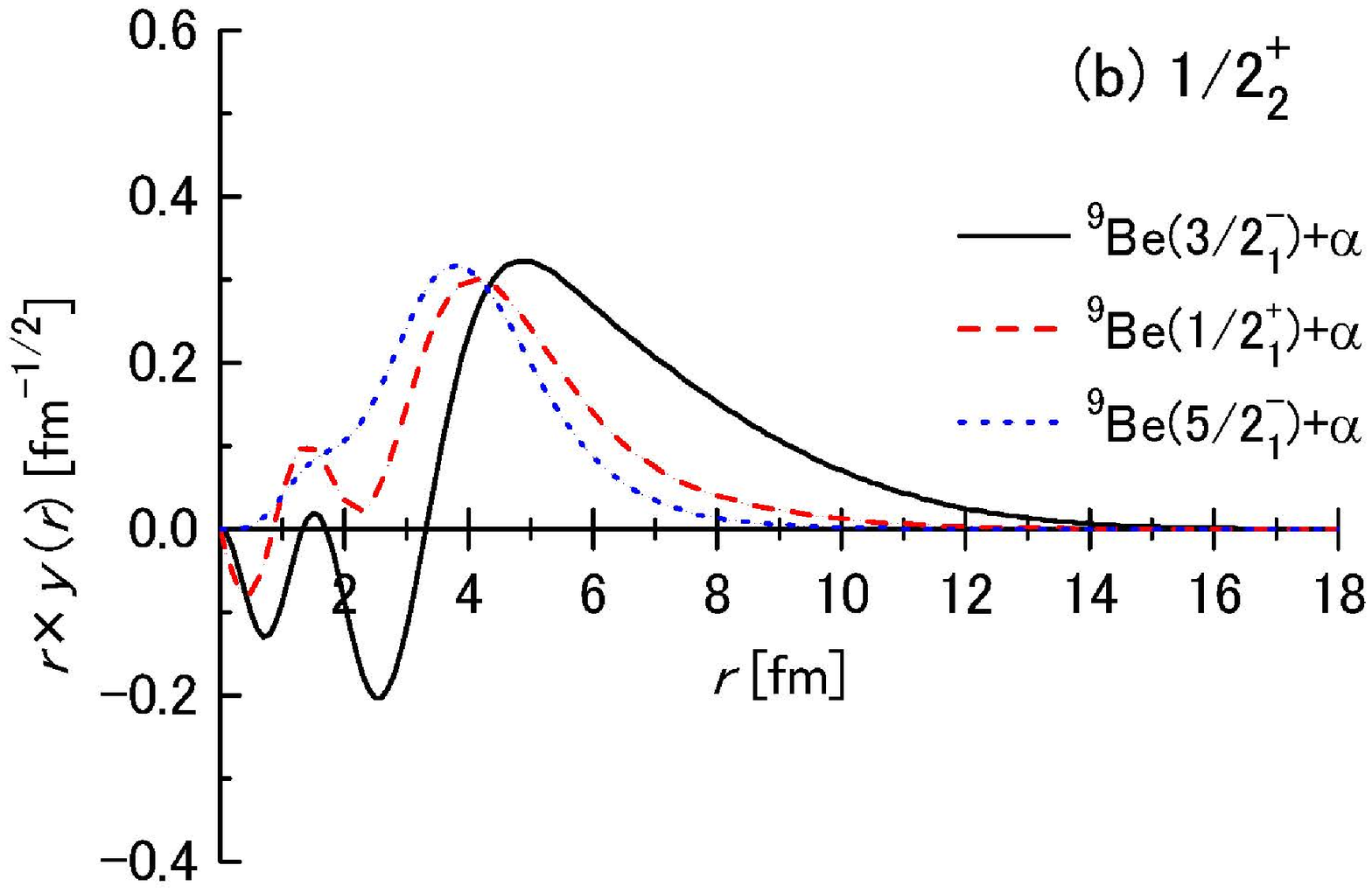}
\\ \vspace*{5mm}
\includegraphics[width=81mm]{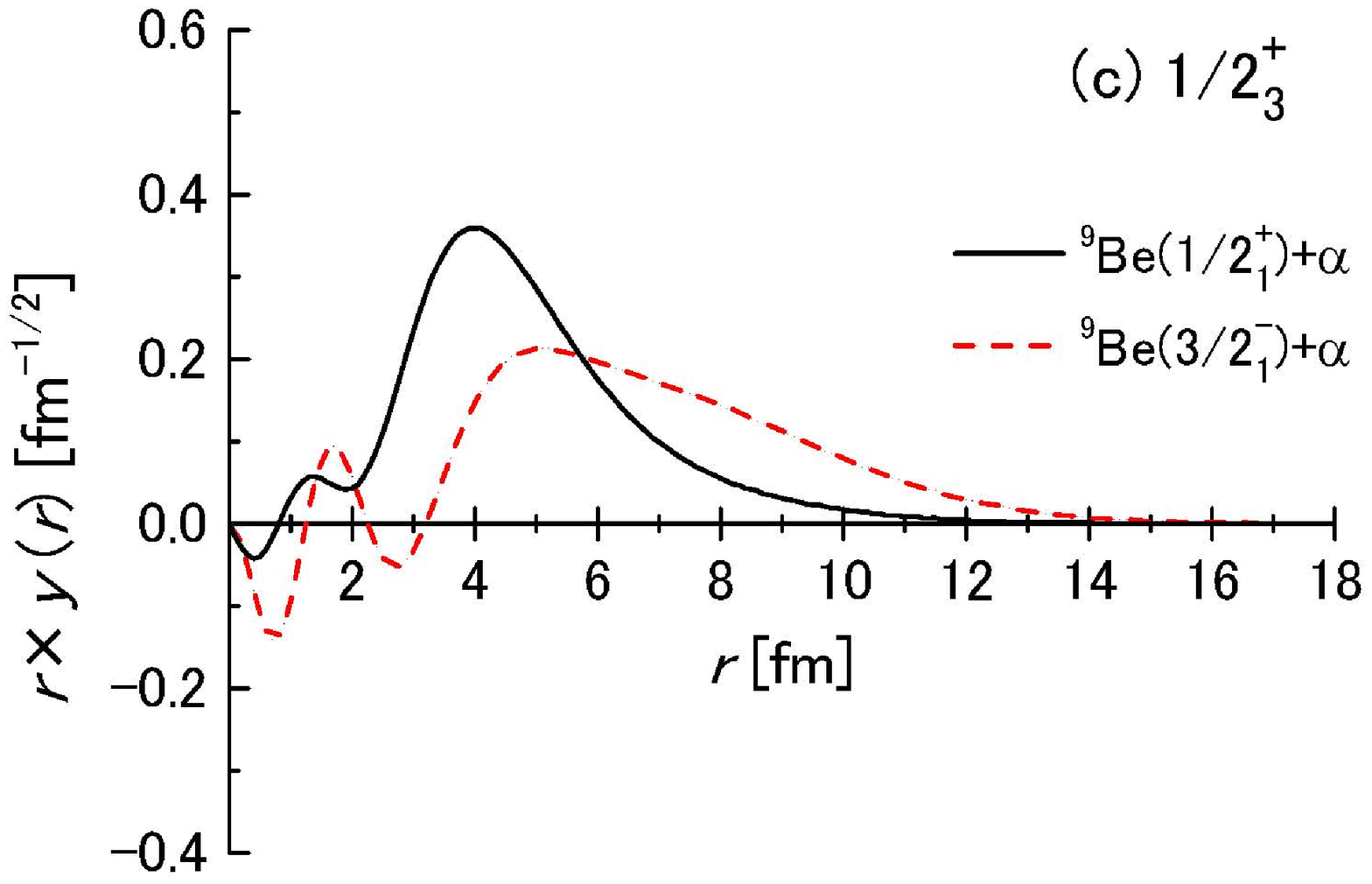}
\includegraphics[width=81mm]{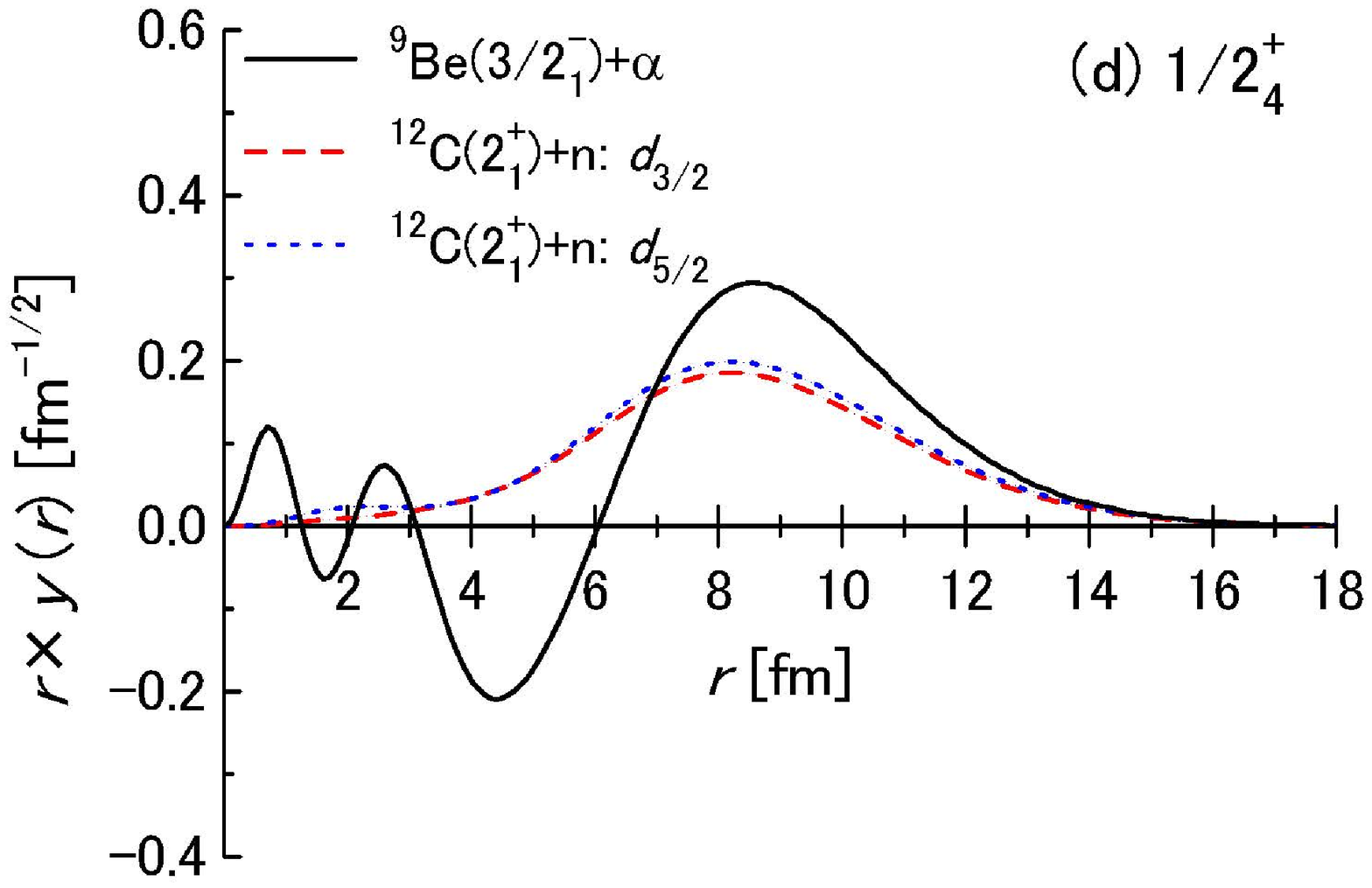}
\\ \vspace*{5mm}
\includegraphics[width=81mm]{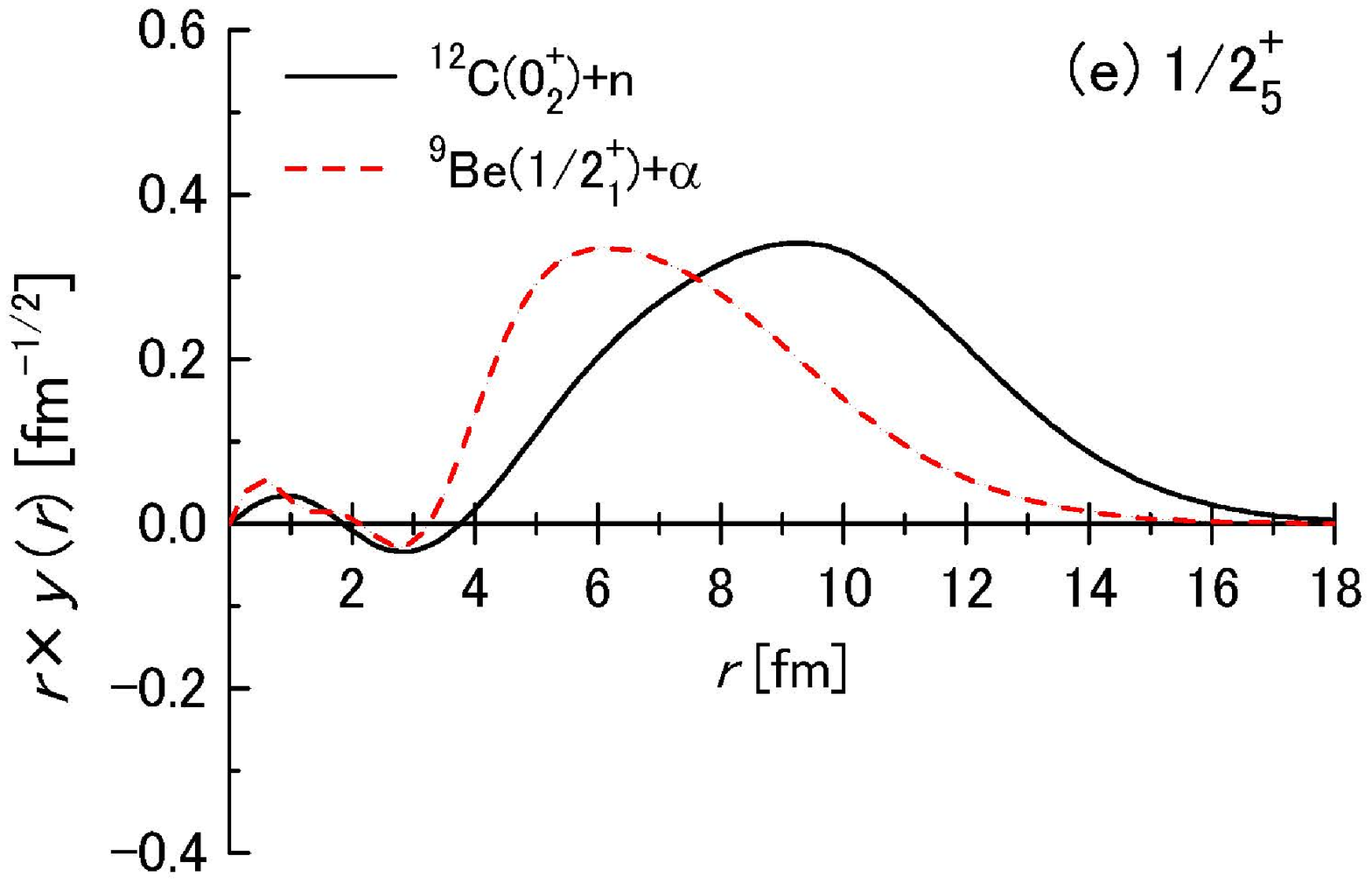}
\caption{
(Color online) Overlap amplitudes of the $^{12}$C+$n$ channels and $^{9}$Be+$\alpha$ channels for the five $1/2^{+}$ states of $^{13}$C defined in Eqs.~(\ref{eq:RWA_neutron}) and (\ref{eq:RWA_alpha}). In the panels we present only the overlap amplitudes with the $S^2$ factor larger that $0.2$ (see Fig.~\ref{fig:s2_factors_12_plus_13c}).
}
\label{fig:rwa_12_plus_13c}
\end{center}
\end{figure}

The calculated energy levels of the $1/2^+$ states are shown in Fig.~\ref{fig:4} together with the experimental ones.
We found the five $1/2^+$ states in the present calculation, although only the three $1/2^+$ states have been observed at the present stage.
The calculated r.m.s.~radii ($R$), spectroscopic factors ($S^2$), and overlap amplitudes are shown in Table~\ref{tab:12plus_ocm}, Fig.~\ref{fig:s2_factors_12_plus_13c}, and Fig.~\ref{fig:rwa_12_plus_13c}, respectively. 

The $1/2^+_1$ state, which is located just below the $^{12}$C($0^+_1$)+$n$ threshold, has the dominant $S^2$ factor of the  $^{12}$C($0^+_1$)+$n$ channel ($S^2\sim 0.65$).
Due to the fact that this state is bound by only $1.9$~MeV with respect to the $^{12}$C($0^+_1$)+$n$ threshold, the overlap amplitude of the $^{12}$C($0^+_1$)+$n$ channel for the $1/2^+_1$ state has a very long tail which is extended up to $r\sim 12$~fm (see Fig.~\ref{fig:rwa_12_plus_13c}(a)).
Thus, the $1/2^+_1$ state has a loosely bound neutron structure, in which the extra neutron moves around the $^{12}$C($0^+_1$) core with $1S$ orbit.
The radius of this state is $R=2.6$~fm, the value of which is by about $10~\%$ larger than that of the ground state ($R$=2.4~fm).
This slightly larger radius is supported by the experimental analysis with the differential cross sections of the inelastic scattering~\cite{ogloblin11}.  
From Fig.~\ref{fig:rwa_12_plus_13c}(a) one notices that the overlap amplitudes of $^9$Be($3/2^-$)+$\alpha$ and $^9$Be($1/2^+$)+$\alpha$ in the $1/2^+_1$ state has the $2P$- and $2S$-like oscillations and their largest peaks are located around $r=3$~fm.   
This fact indicates that the $1/2^+_1$ state has the $^9$Be($3/2^-$)+$\alpha$ and $^9$Be($1/2^+$)+$\alpha$ cluster degrees of freedom.

As for the $1/2^+_{2,3,4,5}$ states, their radii are larger than the ground state and $1/2^+_1$ states (see Table~\ref{tab:12plus_ocm}).
According to the analyses of the $S^2$ factors and overlap amplitudes, the $1/2^+_2$ and $1/2^+_3$ states have, respectively, the dominant cluster structures of $^9$Be($3/2^-$)+$\alpha$ and $^9$Be($1/2^+$)+$\alpha$, in which the $\alpha$ cluster orbits around the $^9$Be($3/2^-$) and $^9$Be($1/2^+$) cores with $3P$ and $3S$ states, respectively, although one sees the non-negligible contributions from the $^{9}$Be($1/2^+,5/2^-$)+$\alpha$ configurations for the $1/2^+_2$ state and $^{9}$Be($3/2^-$)+$\alpha$ ones for the $1/2^+_3$ state, and the $S^2$ factors of the $^{12}$C(Hoyle)+$\alpha$ channels is as small as $S^2=0.10\sim0.15$ for both the $1/2^+_{2,3}$ states.
The overlap amplitudes of the respective dominant channels have the maximum peaks around $r=5$~fm.
As mentioned above, the $1/2^+_1$ state has the $^9$Be($3/2^-$)+$\alpha$ and $^9$Be($1/2^+$)+$\alpha$ cluster degrees of freedom with the $2P$ and $2S$ behaviors in the $^9$Be-$\alpha$ relative motion. 
Thus, the $1/2^+_2$ and $1/2^+_3$ states are regraded as the monopole-type excitation of the relative motions of the $^9$Be($3/2^-$)+$\alpha$ and $^9$Be($1/2^+$)+$\alpha$ cluster degrees of freedom in the $1/2^+_1$ state, respectively, i.e. from $2P$ to $3P$ and $2S$ to $3S$.
This situation is similar to that in the $1/2^-_{2,3}$ states, as discussed in the previous sections.
 In fact the isoscalar monopole matrix elements and ${\rm C0}$ transition matrix elements from the $1/2^+_1$ state to the $1/2^+_{2,3}$ states in Table~\ref{tab:12plus_ocm} are comparable to or larger than those from the $1/2^-_1$ state to the $1/2^-_{2,3}$ states with the $\alpha$ cluster structures (see Table~\ref{tab:12minus_ocm}).
The $1/2^+_3$ state with the dominant $^9$Be($1/2^+$)+$\alpha$ structure is located just above the $3\alpha+n$ threshold.
It is noted that the $1/2^+_1$ state of $^9$Be also appears just above the $2\alpha+n$ threshold (see Fig.~\ref{fig:3}) and this state is pointed out to be a virtual state~\cite{okabe77,arai03}.
Thus, it is interesting to study whether the $1/2^+_3$ state of $^{13}$C turns out to be a virtual state or not with imposing proper boundary conditions for the present four-body $3\alpha+n$ model in near future.

On the other hand, the $1/2^+_4$ state is identified as the $^9$Be($3/2^-$)+$\alpha$ cluster state with the higher nodal behavior, in which the relative wave function between the $^9$Be($3/2^-$) and $\alpha$ clusters has one node more than that in the $1/2^+_2$ state.
Reflecting the characteristic of its higher nodal state, the radius of this state is $R=4.0$~fm, the value of which is larger than that of the $1/2^+_2$ state ($R=3.2$~fm).
  
As for the $1/2^+_5$ state, this state is interpreted as a $3\alpha+n$ gas-like state in $^{13}$C as shown below.
From Figs.~\ref{fig:s2_factors_12_plus_13c}(e) and \ref{fig:rwa_12_plus_13c}(e), one sees that the dominant configuration of the $1/2^+_5$ state is $^{12}$C(Hoyle)+$n$ ($S^2 \sim 0.6$) with an $S$-wave relative motion, strongly coupled with $^9$Be($1/2^+$)+$\alpha$ ($S^2\sim 0.5$).
The characteristic features of this state are that 1)~the nodal behavior of the overlap amplitudes of the $^{12}$C(Hoyle)+$n$ channel together with the $^9$Be($1/2^+$)+$\alpha$ one almost disappears in the inner region ($r < 3$~fm), indicating that the Pauli-blocking effect is significantly reduced, 2)~the relative orbital angular momentum in the $^{12}$C(Hoyle)+$n$ channel as well as the $^9$Be($1/2^+$)+$\alpha$ one is $S$-wave, and thus all the $\alpha-\alpha$ relative motions as well as the $\alpha-n$ ones in this state are dominantly $S$-wave, and 3) the radius of this state is $R=4.3$~fm, the value of which is similar to the calculated result of the Hoyle state with the $3\alpha$ OCM.
These results indicate that the $1/2^+_5$ state has a $3\alpha+n$ gas-like structure.

\begin{figure}[t]
\begin{center}
\includegraphics*[width=8.6cm]{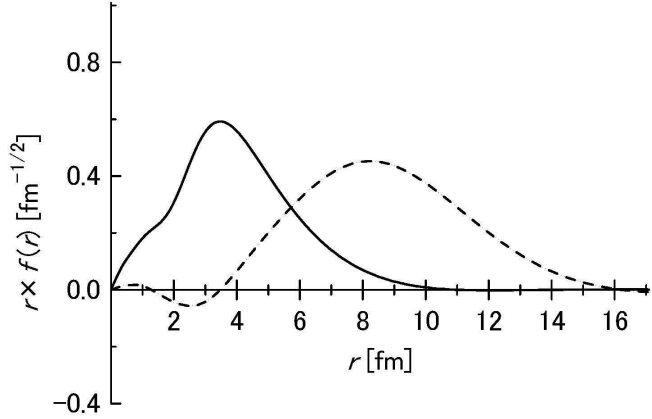}
\caption{
Radial behaviors of the dominant $S$-wave single-$\alpha$ orbit (real line) with the occupation probability of $48~\%$ and the dominant $S$-wave extra-neutron orbit, $(S)_n$,  (dashed line) with the occupation probability of $65~\%$ in the $1/2^+_5$ state, which are obtained by diagonalizing the single-$\alpha$-cluster density matrix and single nucleon density matrix of the extra neutron defined in Eqs.~(\ref{eq:density_alpha}) and (\ref{eq:single_nucleon_density}), respectively. 
See the text for the details.
}
\label{fig:single_orbits}
\end{center}
\end{figure}

The nature of the $\alpha$-particle condensate in the $1/2^+_5$ state can be investigated with the analysis of the single-$\alpha$ orbit and occupation probability obtained by diagonalizing the single-$\alpha$ cluster density matrix defined in Eq.~(\ref{eq:density_alpha}).
It is reminded that in the Hoyle state, typical of $\alpha$ condensate state, the occupation probability of the $0S$-wave single-$\alpha$ orbit (zero-node $S$-wave Gaussian type) in its state amounts to be $70~\%$, and thus the Hoyle state is described by the product states of $\alpha$ clusters, $(0S)^3_\alpha$, with the probability of $70~\%$~\cite{yamada05}.  
We found that the dominant $S$-wave single-$\alpha$ wave function in the $1/2^+_5$ state is the $0S$-type (see Fig.~\ref{fig:single_orbits}), and the occupation probability of this orbit amounts to be $48~\%$.
The radial behavior shown in Fig.~\ref{fig:single_orbits} is similar to that of the single-$\alpha$-cluster $S$-wave orbit in the Hoyle state (see Fig.~4(a) in Ref.~\cite{yamada05}). 
The total occupation probability of the $S$ wave, i.e., the total sum from $0S$ and $1S$ etc., amounts to be $57~\%$, whereas that of the $P$ wave stands at $14~\%$.
On the other hand, we found that the extra neutron orbit with the largest occupation probability obtained by diagonalizing the single-particle density matrix for the extra neutron defined in Eq.~(\ref{eq:single_nucleon_density}) is $S$-wave and the occupation probability amounts to be as large as $65~\%$ (hereafter we call this orbit $(S)_n$).
The radial behavior of the $(S)_n$ orbit is shown in Fig.~\ref{fig:single_orbits}, where its nodal behavior almost disappears in the inner region ($r<3$~fm), indicating significantly reduced Pauli-blocking effect.
These results indicate that the $1/2^+_5$ state can be regarded as an $\alpha$-particle condensate-like state, $(0S)^3_{\alpha}(S)_n$, with the probability of $2.09/4\sim52~\%$, where the dominator denotes the particle number (three $\alpha$-clusters and one extra neutron) and the numerator is the particle number occupied in the $(0S)^3_\alpha(S)_n$ configuration, that is, $3\times0.48+0.65$.
From Fig.~\ref{fig:single_orbits} we note that the extra neutron moves at the outer side of the $3\alpha$-gas-like region, avoiding the Pauli-blocking effect between the extra neutron and $3\alpha$ clusters.
 
The reasons why the occupation probability of the single-$\alpha$-cluster $0S$ orbit in the $1/2^+_5$ state ($48~\%$) is smaller than that in the Hoyle state ($70~\%$) are given as the follows:~The $P$-wave $\alpha$-$n$ force is attractive to produce a bound state and resonant states in the $2\alpha+n$ system.
This means that the $^9$Be($3/2^{-}$,$1/2^{-}$)+$\alpha$ correlations are enhanced in the $3\alpha+n$ system, as discussed in the previous sections (see Sec.~\ref{subsec:structures_12minus}).
In the $1/2^+_5$ state the $^9$Be($3/2^{-}$,$1/2^{-}$)+$\alpha$ correlations with $P$-wave relative motion can be seen, for example, in the calculated results of the $S^2$ factors (see Fig.~\ref{fig:s2_factors_12_plus_13c}(e)) as well as the occupation probability of the $P$-wave single-$\alpha$ orbit being $14~\%$ as mentioned above.
This two-body correlations are considered to hinder the growth of the $\alpha$-particle-condensation aspect in the $^{13}$C system.

\textcolor{black}{
Here we present some remarks on positive parity states, in particular, higher spin states.
The shell model calculation by Millener et al.~\cite{millener89} indicates that the $7/2^{+}_{2}$, $5/2^{+}_{4}$, and $3/2^{+}_{3}$ states (including a $1/2^+$ state) are dominantly formed by coupling an {\it sd} nucleon to $^{12}$C states with the $[431]$ spatial symmetry and the former three high-spin states are strongly excited by the C3 transitions (see Table~II and Fig.~9 in Ref.~\cite{millener89}), although several positive parity states are also suggested to be produced by coupling a nucleon to the $[44]$ symmetry states of $^{12}$C.
The group of the former states can not be addressed in the present $3\alpha+n$ cluster model.
However, an extended cluster model, $\alpha+\alpha+"3N"+"N"+n$, can be addressed to study the structure of the high-spin states formed by coupling an $sd$ nucleon to $^{12}$C states with the $[431]$ symmetry, where $"3N"$ ($"N"$) denotes $^3$H-$^3$He clusters (nucleon), because the model space of $\alpha+\alpha+"3N"+"N"$ has the $[431]$ symmetry states of $^{12}$C as well as the [44] ones.
In the hypernuclear physics, an extended cluster model, $\alpha+"3N"+"N"+\Lambda$, has been applied to the study of the structure of $^9_\Lambda$Be hypernucleus up to $E_x \sim 25$~MeV~\cite{yamada88,hiyama09}, where the high-lying (low-lying) states of $^9_\Lambda$Be are dominantly formed by coupling a $\Lambda$ particle to the $[31]$ ($[4]$) symmetry states of $^8$Be.
This hypernuclear extended cluster model has succeeded in describing the structures of $^9_\Lambda$Be up to $E_x \sim 25$~MeV and also reproducing the excitation functions of the $^9$Be({\rm in-flight}~$K^-,\pi^-)$,  $^9$Be$(\pi^+,K^+)$, and $^9$Be$({\rm stopped}~K^-,\pi^-)$ reactions up to $E_x \sim 25$~MeV, where the single-nucleon parentage to the $[31]$ symmetry states (including the $[4]$ ones) of $^{8}$Be from the ground state of $^{9}$Be are addressed within the cluster model~\cite{yamada88,hiyama09}.
Thus, the extended cluster model of $^{13}$C is promising and is one of our future subjects.      
}

\section{Summary}\label{sec:summary}

We have investigated the structure of the $1/2^{\pm}$ states of $^{13}$C up to around $E_x \sim 16$~MeV with the full four-body $3\alpha$+$n$ OCM.
The model space describes nicely the structure of the low-lying states ($0^+_{1,2}$, $2^+_{1}$, $4^+_{1}$, $3^-$, and $1^-$) of $^{12}$C, including the $2^+_2$, $0^+_3$, $0^+_4$ states above the Hoyle state, with the $3\alpha$ OCM, together with those of $^9$Be, $^8$Be, and $^5$He with the $2\alpha$+$n$, $2\alpha$, and $\alpha$+$n$ OCM\rq{}s, respectively.
We have succeeded in reproducing all the five $1/2^{-}$ states and three $1/2^{+}$ states  observed up to $E_x \sim 16$~MeV.
It was  found that the $1/2^-_{2}$ and $1/2^-_{3}$ states have mainly $^9$Be($3/2^-$)+$\alpha$ and $^9$Be($1/2^-$)+$\alpha$ cluster structures, respectively, while the ground state $1/2^-_1$ has a shell-model-like structure. 
The $1/2^-_{4}$ and $1/2^-_{5}$ states are characterized by the dominant structures of $^9$Be($3/2^-$)+$\alpha$ and $^9$Be($1/2^-$)+$\alpha$ with higher nodal behaviors, respectively.

The present calculations \textcolor{black}{for the first time have provided the reasonable agreement with} the experimental data on the ${\rm C0}$ matrix elements ${\mathcal M}({\rm C0})$ of the $1/2^{-}_{2}$ ($E_x=8.86$~MeV) and $1/2^-_3$ ($E_x=11.08$~MeV) states obtained by the $(e,e')$ reaction, isoscalar monopole matrix elements ${\mathcal M}({\rm IS})$ of the $1/2^{-}_{2}$ ($E_x=8.86$~MeV), $1/2^{-}_{3}$ ($E_x=11.08$~MeV), and $1/2^-_4$ ($E_x=12.5$~MeV) states by the $(\alpha,\alpha')$ reaction, and r.m.s.~radius of the ground state.
It is noted that the experimental values of ${\mathcal M}({\rm C0})$ and ${\mathcal M}({\rm IS})$ are strong as to be comparable to the single particle strengths.
The reason why the $^9$Be+$\alpha$ cluster states are populated by the isoscalar monopole transition and ${\rm C0}$ transition from the shell-model-like ground state has been discussed in detail.
We found that this mechanism, which is common to those in $^{16}$O, $^{12}$C, $^{11}$B, and $^{12}$Be etc., originates from the dual nature of the ground state~\cite{yamada08_monopole,yamada10}:~The ground state in light nuclei have in general both the mean-field degree of freedom and cluster degree of freedom, the latter of which is activated by the monopole operator and then cluster states are excited from the ground state. 
\textcolor{black}{The} present results indicate that the $\alpha$ cluster picture is inevitable to understand the low-lying structure of $^{13}$C, and the {C0} transitions together with the isoscalar monopole transitions are also useful to explore cluster states in light nuclei.

From the analyses of the spectroscopic factors and overlap amplitudes of the $^{9}$Be+$\alpha$ and $^{12}$C+$n$ channels in the $1/2^-$ states, dominant $^{12}$C(Hoyle)+$n$ states do not appear in the $1/2^-$ states in the present study.
This is mainly due to the effect of the enhanced $^9$Be+$\alpha$ correlation induced by the attractive odd-wave $\alpha$-$n$ force:~When an extra neutron is added into the Hoyle state, the attractive odd-wave $\alpha$-$n$ force reduces the size of the Hoyle state with the $3\alpha$ gas-like structure and then $^9$Be+$\alpha$ correlation is significantly enhanced in the $3\alpha$+$n$ system.
Consequently the $^9$Be($3/2^-$)+$\alpha$ and $^9$Be($1/2^-$)+$\alpha$ states come out as the excited states, $1/2^-_{2}$ and $1/2^-_3$, respectively.
On the other hand, higher nodal states of the $1/2^-_{2,3}$ states, in which the $^9$Be-$\alpha$ relative wave function has one node higher than that of the $1/2^-_{2,3}$ states, emerge as the $1/2^-_{4}$ and $1/2^-_5$ states, respectively, in the present study.
It is reminded that the $0^+_3$ state of $^{12}$C has $^8$Be+$\alpha$ structure with higher nodal behavior. 
Thus, the $^9$Be+$\alpha$ cluster states with higher nodal behavior,  $1/2^-_{4,5}$, are regarded as the counterpart of the $0^+_3$ state in $^{12}$C.

As for the $1/2^+$ states, the $1/2^+_1$ state appears as a bound state by $1.9$~MeV below the $^{12}$C($0^+_1$)+$n$ threshold.
This state dominantly has a loosely bound neutron structure, in which the extra neutron moves around the $^{12}$C($0^+_1$) core with $1S$ orbit.
The calculated radius of the $1/2^+_1$ state ($R=2.6$~fm), slightly lager than that of the ground state ($R=2.4$~fm), is consistent with the experimental suggestion~\cite{ogloblin11}.
It was found that the $1/2^+_2$ and $1/2^+_3$ states have mainly the $^9$Be($3/2^-$)+$\alpha$ and $^9$Be($1/2^+$)+$\alpha$ structures, respectively, and their radii are around $R=3$~fm.
These two states are characterized by the strong isoscalar monopole excitations from the $1/2^+_1$ state.
We have discussed in detail the mechanism of why the two cluster states are excited by the monopole transitions from the $1/2^+_1$ state.
On the other hand, we found that the $1/2^+_4$ and $1/2^+_5$ states have dominantly the $^9$Be($3/2^-$)+$\alpha$ structure with higher nodal behavior and $3\alpha$+$n$ gas-like structure, respectively, although experimentally the two states have not been identified so far. 
The $1/2^+_5$ state with a larger radius ($R\sim 4$~fm) is described by the product states of constituent clusters, having a configuration of $(0S)^3_\alpha(S)_n$, with the probability of $52~\%$.
Thus, the $1/2^+_5$ state can be regarded as an $\alpha$-condensate-like state.

\textcolor{black}{
It is interesting to study the structure of the higher angular momentum states ($J^{\pi}=3/2^{\pm}$, $5/2^{\pm}$, $\cdots$) of $^{13}$C with the present $3\alpha+n$ cluster model.
In addition it is also important to investigate the structure of the $^{13}$N mirror nucleus with $3\alpha+p$ cluster model. 
The Coulomb energy shifts between $^{13}$C and $^{13}$N as well as the decay widths of their excited states and M1 transition strengths etc.~will offer another interesting information on their structures.   
These theoretical studies are now in progress.
}
The results will be given elsewhere.

\section*{Acknowledgments}

The authors would express thanks to Profs.~H.~Horiuchi, A.~Tohsaki, P.~Schuck, and  G.~R\"opke for many useful discussions and comments.
This work was partially performed with the financial support by JSPS KAKENHI Grant Number 26400283 (T.~Y.), and by HPCI Strategic Program of Japanese MEXT, JSPS KAKENHI Grant Number 25400288, and RIKEN Incentive Research Projects (Y.~F.).


\end{document}